\documentclass[letterpaper,12pt]{article}
\usepackage[utf8]{inputenc}
\usepackage{lipsum}
\usepackage[margin=0.7in]{geometry}
\usepackage{mathptmx}
\usepackage{xfrac}
\usepackage{graphicx}
\usepackage{float}
\usepackage{subcaption} 
\usepackage{wrapfig}
\usepackage{amsmath}
\usepackage{amssymb}
\usepackage{epstopdf}
\usepackage{setspace}
\usepackage{mathtools, nccmath}
\usepackage{fancyhdr}
\usepackage[
    style=numeric-comp,
    bibstyle=phys,
   articletitle=true,
    biblabel=brackets,
    backend=biber,eprint=true
]{biblatex}
\usepackage{hyperref}
\hypersetup{
  colorlinks = true,
            allcolors  = red,
            citecolor  = blue}
\usepackage[compact]{titlesec}
    \titlespacing{\section}{0pt}{1ex}{1ex}
    \titlespacing{\subsection}{0pt}{1ex}{0ex}
    \titlespacing{\subsubsection}{0pt}{0.5ex}{0ex}
\usepackage{soul}
\DeclareUnicodeCharacter{0327}{,}
    
\def\bea{\begin{eqnarray}}
\def\eea{\end{eqnarray}}
\def\be{\begin{equation}}
\def\ee{\end{equation}}
\def\nn{\nonumber}

\def\om{\omega}

\def\t{\tau}

\def\a{\alpha}

\def\p{\partial}

\title{ Interaction of Gravitational Waves with\\ Yang-Mills fields}
\author{Narasimha Reddy Gosala \footnote{narasimha.gosala@uleth.ca}, Arundhati Dasgupta \footnote{arundhati.dasgupta@uleth.ca}\\ 4401 University Drive, University of Lethbridge, Lethbridge, Canada}
\date{\vspace{-5ex}}

\begin{document}

\maketitle

\begin{abstract}
    In this paper, we discuss the interaction of non-Abelian SU(2) Yang-Mills progressive waves with gravitational waves. We solve and obtain some interesting solutions to pure Yang-Mills equations in different backgrounds, and perturbative solutions induced due to gravitational waves. These perturbations show `beat patterns' and depending on boundary conditions, changes in frequency. In flat space-time, when the Yang-Mills fields and the gravitational waves are in the same direction there is no interaction, unless there is self interaction of the Yang-Mills fields. In the system with non-zero self interaction the amplitudes of the perturbation are inversely proportional to the Yang-Mills coupling constant. In a cosmological background, the Yang-Mills fields and the gravitational wave interact when they are in the same direction even without self interaction of the Yang-Mills progressive fields. We find that in the electroweak symmetry broken phase of the gauge fields, the interactions are perturbative only for an infinitesimal time.
\end{abstract}

\section{Introduction}

Since the discovery of gravitational waves (GW), there have been new developments in the field of astronomy in the understanding of binary systems of black holes and neutron stars. GWs provide a way to probe the very early universe since GWs get decoupled from the matter just after their production \cite{GW1, GW2Allen}. The strongly interacting matter (strong nuclear force) is described by the non-Abelian gauge theories having SU(3) symmetry. At very high temperatures, above the critical temperature (Hagedorn temperature $ \sim $ 150 MeV), it is well known that the interaction between quarks and gluons is very weak, resulting in a de-confined phase known as Quark Gluon Plasma (QGP) \cite{qgpa}. The temperatures of this phase in the very early universe persist for around $10^{-6}$ sec. During this phase, it may be expected that gluons propagate. In the chronology of the universe, before the QGP phase, there are many other phenomena like inflation, preheating, and Electro Weak phase transition, that can produce GWs, termed Primordial GWs (PGW). It may be expected that the PGWs interact with the QGP. In the post-merger data analysis of a recent neutron star merger, GW interacted with quark-gluon matter \cite{qgpm}. Motivated by this, and as the search for PGW is still nascent, we investigate the interaction of YM fields with gravitational waves in general backgrounds. We study the interaction of SU(2) YM waves with the gravitational waves in flat space background and de-Sitter background. These are classical calculations, and we use the previous work of \cite{COLEMAN,OT}, for the YM fields. The classical YM equations were solved previously in \cite{Actor,LO,BLAIZOT,CAMPBELL,COLEMAN,BASLER,rabinowitch2006,rabinowitch2007,rabinowitch2008,RACZKA,Tsapalis,Baseian} with propagating modes, and in terms of non-abelian plane waves. For the solitons and instanton solutions of the same systems, see \cite{solit}. A study of how GW interact with these is work for the future. A discussion on how YM condensates interact with GW can be found in \cite{condensate}. Note that in the study of YM progressive waves in backgrounds with gravitational waves, our calculations are \textit{completely new}. Given the new emergence of gravitational wave detection, these calculations are important for further use in gravitational wave research.  
For new work in the study of YM energy-momentum tensor as the source of gravitational waves, see \cite{symh}. However, the point of view we take here uses the following scenario: The gravitational wave is generated by some other process, and it interacts with a YM field. These YM fields are classical solutions of the YM Lagrangian, without sources in the given region. We allow for these progressive waves to be generated in experiments, and could be in controlled collimated beams from future technology, or these could be waves in some early universe cosmic soup. Our calculations are relevant for the symmetry broken massive W, Z, bosons observed in colliders. In the early universe of course, one studies thermal processes. However, the interaction has to be identified and this paper provides the theoretical basis of YM fields and GW wave interactions. In fact, the study of GW waves with matter fields is relevant for
 the (i) early universe qgp, (ii) colliders, (iii) neutron star mergers, and (iv) new ways of GW detection. 
For the YM fields, the focus of this paper, we amplify the motivations further. \\
(i) Early Universe:  The interaction of YM fields and the gravitational wave can be modelled using the propagation of YM waves in de-Sitter space as we show here. De-Sitter cosmology describes inflation. Our results can be extended to other cosmologies, using different formulas for the scale factor.\\
(ii) In heavy ion collisions, de-confined phases of the gluons are seen for a very short time. However, a GW passing through the collider might affect the trajectory of the YM massive waves. We perform a calculation of the YM fields coupled to Higgs Boson in presence of a GW to that effect.\\
(iii) In neutron star mergers, GWs are released. In \cite{qgpm}, it is shown that these mergers can shed light on the interaction of GW with quark matter and quark-gluon phase transitions. These processes like multi-messenger astronomy and post-merger scenarios of GW emission are also therefore relevant motivation for the current paper.\\
(iv) New ways of GW detection: We suggest for e.g. the use of electro-weak transition, which is otherwise studied without gravitational waves, in tandem with a GW event. The transition ``rates'' will be modified due to the event happening in presence of a GW. The calculations in this paper for the propagation of massive $W, Z$ Bosons in presence of a GW, therefore, acquire relevance.\\ 

In this paper, we try to find how the \textit{SU(2) YM progressive wave} gets modified in the presence of a gravitational wave, just like the Electromagnetic (EM) waves, scalars and neutrinos in \cite{Akash} \& \cite{Morales} respectively.  These SU(2) YM fields represent a W and Z Boson field or a gluon field. We discuss the different types of SU(2) YM waves depending on the nature of the solutions, and their dependence on the coupling constant ($ g_{\rm{YM}} $).  We find that there is no interaction between GW and  Type I SU(2) YM waves when they are propagating in the same direction in Minkowski spacetime, similar to the electromagnetic waves as in \cite{Akash}. We did find an interaction between the waves in the Type II case where the nonlinear terms for the YM equations were considered, or self-interaction terms are non-zero. We also analyze the perturbed solutions when the waves are anti-parallel or perpendicular to each other. The equations we solve are inhomogeneous differential equations, and depending on the boundary condition, we find exact forms for the perturbations induced in the YM fields due to the gravitational wave. These perturbations can be detected as changes in 
the YM field energy-momentum tensor. We provide spectro-graphic analysis and plots of these changes.


In this paper, we also find that if we study the interactions of the W, Z bosons in the symmetry-broken phase with the GWs, the perturbation grows beyond the "perturbative regime" very quickly. The interactions are cubic in the fields, and the magnitude of the `source terms' for the perturbations increases, despite the weak amplitude of the gravitational waves. We find that the perturbation calculation is justified, as the W, Z bosons have a half-life of about $10^{-25} {\rm s}$ and the effect of the gravitational wave is encoded in a perturbative form in the YM field for that short time. 

In the next section, we review the different types of YM field solutions \cite{COLEMAN,OT} as progressive waves. In the third section, we discuss the interaction of YM waves with gravitational waves in a flat background. We take the waves in different orientations like when they are perpendicular to each other, and when they are in the same or opposite direction. In the fourth section, we discuss the interaction of YM waves with gravitational waves in the de-Sitter background. We discuss the implications of this work in the view of early universe cosmology. As in the current epoch the SU(2) symmetry is broken, we discuss the YM fields which acquire mass in presence of a Higgs vacuum expectation value (vev). The interaction of the GWs is studied with these massive YM fields in the fifth section. The last section is the conclusion and description of future open problems.

\section{Yang-Mills Plane waves}{\label{sec2}}
The correct description of the gluon field requires an understanding of SU(3) YM theory. But the basic characteristics can be described by classical SU(2) YM theory as SU(2) is a subgroup of SU(3). We will consider the analysis of SU(2) fields in this paper because SU(2) is a subgroup of SU(3) and also it is the symmetry group for weak interactions.  The Lagrangian density can be written as 
\begin{equation}
    \mathcal{L} = - \frac{1}{4} F^{a\mu\nu} F^a_{\mu\nu}
\end{equation}
where $F^a_{\mu\nu}$ is the anti-symmetric field strength tensor of gauge field $A^a_\mu $ given by 
\begin{equation}{\label{eqF}}
    F^a_{\mu\nu} = \partial_\mu A^a_\nu - \partial_\nu A^a_\mu + g_{\rm{YM}} \epsilon^{abc} A^b_\mu A^c_\nu
\end{equation}
where $g_{\rm{YM}}$ is the YM coupling constant and $ F_{\mu\nu} = F^{a}_{\mu\nu} T^a, $ with the generators $T^a$ of Lie algebra, indexed by $a$, satisfying $ {\rm Tr}(T^a T^b) = \frac{1}{2} \delta^{ab} ,\  [T^a,T^b] = i \epsilon^{abc} T^c $ here $\epsilon^{abc}$ are the structure constants of Lie algebra. Greek indices represent the space-time coordinates and Latin indices represent the internal indices of the gauge group, for a SU(N) group, one has $a,b,c = 1,., N^2-1$.

The SU(2) YM field equations are 
\begin{equation}{\label{eqYM}}
    \nabla_\mu F^{a \mu\nu} + g_{\rm{YM}} \  \epsilon^{abc}\  A_{\mu}^{b}\  F^{c\mu\nu} = 0,
\end{equation}
where $\nabla_{\mu}$ is the covariant derivative. Wave solutions for the YM equation in flat space-time were first discovered by Sidney Coleman \cite{COLEMAN}.


In order to exploit the full behaviour of YM waves, we discuss the two families of YM waves as suggested in \cite{OT}. One was labelled as Type I and another was labelled as Type II. Here the difference between the two families is due to the nature of the commutation term or self-interaction term in the Lagrangian.

\subsection{Type I waves}{\label{sec2:typeI}}
The type I waves are characterized based on the following conditions as discussed in papers \cite{CAMPBELL} \& \cite{OT} :
$ [A_\mu,A_\nu] = [A_\mu, F^{\mu\nu}] = 0 $ and $ [A_\mu,F^{\nu \rho}] \neq 0 $ which says that the energy and momentum densities are equal in magnitude.

The ansatz for the type I waves as given in \cite{OT} is
\begin{equation}
A^a_\mu = (\psi_\alpha f^a_\alpha (U) + h^a(U) ) \  \partial_\mu U,
\end{equation} 
here U and $\psi_\alpha $ are functions of $x_\mu$, $f^a_\alpha $ and $h^a$ depend on U only and the index $\alpha=1..4$. The function $h^a(U)$ can be gauge transformed away. The functions $\psi_{\alpha}$ represent degrees of freedom which is an arbitrariness in the form of the ansatz \cite{OT}.

The equation of motion for this form of the YM fields using the flat space-time metric with the additional conditions of $\p^{\mu}U \p_{\mu} U=0$ and $\p^{\mu}U\p_{\mu}\psi_{\alpha}=0$ is
\begin{equation}{\label{eq:2.1eom}}
 \p^{\mu}\p^\nu U \p_\mu \psi_\alpha + \p^\nu U \Box \psi_\alpha -    \p^{\lambda}\p^{\nu} \psi_{\alpha} \p_{\lambda} U - \p^{\nu}\psi_{\alpha} \Box \ U =0,
\end{equation}
where $\Box = \partial^{\mu}\partial_{\mu}$ is the d'Alembertian operator. One can see that the equations of motion and conditions do not involve $f^a_\alpha$ which has an arbitrary functional nature. This is facilitated by the fact that any terms containing derivative of the form $\partial_{\mu}f^a_\alpha(U)= \partial_U f^a_\alpha(U)\partial_{\mu}U$ are eliminated by the above conditions.

We take two different forms of the function $\psi_{\alpha}$ yielding Type Ia and Type Ib waves.
\subsubsection{Type Ia waves}{\label{sec2:typeIa }}
If we choose $\psi_1=x$ and the other values of $\psi_{\alpha}, \alpha=2,3,4$ are set to zero, then one gets the following equation for the unknown $U$ function 
\begin{equation}
  \Box \ U=0  ,
\end{equation}
with the conditions $\p_1 U=0$ and $\p_\mu U \p^\mu U = 0$ \cite{OT}.

In the above choice of $\psi_{\alpha}$ is arbitrary, and we have taken the easiest option, to show the resultant equation of motion. In the next we discuss another choice.
\subsubsection{Type Ib waves}{\label{sec2:typeIb}}
As the choice of $\psi_{\alpha}$ is arbitrary, subject to the conditions specified above, we also use a more general form for $\psi_{\alpha}$ such that $\partial_\mu \partial^\mu \psi_\alpha = 0 $, and $\psi_\alpha = A_\alpha e^{i\textbf{q}\cdot \textbf{x}}$ where $q_{\mu}$ is a real four-vector. In general, the function can be a superposition of the Fourier modes. Then, Eq. (\ref{eq:2.1eom}) reduces to 
$$ \Box\ {U} = 0 ,$$
with the conditions $q_\mu \partial^\mu U = 0$ and and $\p_\mu U \p^\mu U = 0$. Note this choice is considered new in the paper and was not discussed in \cite{OT}.

\subsubsection{Coleman waves}{\label{sec2:coleman}}
The Coleman solution \cite{COLEMAN} which was obtained as plane-progressive YM waves, is a special case of type I wave solutions. If we take $U(x)  = t \pm z $,  $\psi_1(x)  = x $,  $ \psi_2(x)  = y$ , $ \psi_3(x)  = 0 $, $ \psi_4(x)  = 0$,
we get the Coleman solution:
\begin{align}
    A^a_t  = \pm A^a_z & = x \ f^a_1(t\pm z) + y \ f^a_2(t\pm z), &     A^a_x \ = A^a_y \  = 0.
\end{align}

\subsection{Type II waves}{\label{sec2:typeII}}
In order to understand the full nature of non-linearity, we consider the type II waves which satisfy the following criteria: $ [A_\mu,A_\nu] \neq 0, [A_\mu, F^{\mu\nu}] \neq 0 $ and $[A_\mu,F^{\nu \rho}] \neq 0$. We consider the following ansatz as given in \cite{OT}

\begin{equation}{\label{ansatz2}}
    A^a_\mu = \delta^a_1  \ \Phi(u) \  k_\mu + \delta^a_3 \ \Psi(u) \ l_\mu, 
\end{equation}
where $ u= p_\mu x^\mu + e; $
$e$ is a arbitrary constant and the three vectors $k_\mu,l_\mu,p_{\mu}$ are orthogonal to each other.
By substituting the above ansatz in YM equations, we get the following equations 
\begin{equation}{\label{eq:typeIIeq1}}
 \partial_\mu \partial^\mu \Phi - g_{\rm{YM}}^2 \ l^2\  \Phi  \Psi^2  = 0 
 \end{equation}
 \begin{equation}{\label{eq:typeIIeq2}}
\partial_\mu \partial^\mu \Psi - g_{\rm{YM}}^2 \  k^2\  \Psi  \Phi^2  = 0.
 \end{equation}
 As the functions ($\Phi$ and $\Psi$) are chosen to depend only on single variable $u$ as in \cite{OT}, and motivated by the EM plane wave solutions. We can rewrite the above equations as 
 \begin{align}{\label{eqans21}}
         \Phi''(u) - \frac{g_{\rm{YM}}^2 \ l^2}{p^2}\  \Phi(u)  \Psi(u)^2 & = 0 \\
         \label{eqans22}
         \Psi''(u) - \frac{g_{\rm{YM}}^2 \  k^2}{p^2}\  \Psi(u)  \Phi(u)^2  & = 0.
    \end{align}
These are two coupled nonlinear differential equations. The solutions of Eqs. (\ref{eqans21}\&\ref{eqans22}) can be written in terms of Jacobi elliptic functions.
\begin{align}{\label{eq:typeIIsol}}
        \Phi(u) & = \pm \frac{1}{\sqrt{\beta}} {\rm cn}(u,\frac{1}{2})\\
         \Psi(u) & = \pm \frac{1}{\sqrt{\alpha}} {\rm cn}(u,\frac{1}{2})
\end{align}
where $\alpha = \frac{-g_{\rm{YM}}^2 l^2}{p^2}$,$\beta = \frac{-g_{\rm{YM}}^2 k^2}{p^2}$ and $\rm {cn}(u,\lambda^2)$ is the cosine Jacobi elliptic  function with elliptic modulus $\lambda$. These solutions correspond to wave solutions only if $p_\mu$ is a timelike vector. In the next section, we see how the GW interacts with the above YM propagating solutions.

\section{Interaction of Yang-Mills waves with Gravitational waves in Minkowski Spacetime }{\label{sec3}}

We study the interaction of a GW travelling in the $z$-direction and a YM wave. This calculation is new, and we use perturbation techniques. However, due to the nature of the nonlinear coupling of YM waves, we find that the YM wave gets non-trivially perturbed.
We consider a gravitational wave $h_{\mu\nu}$ with only one polarisation $h_+$ propagating at frequency $\omega_g$ with an amplitude of $A_+$ in the $z$-direction. The gravitational wave metric is taken as \cite{GW1}
\begin{equation}
     ds^2 = -dt^2 + (1 + h_+(t,z))dx^2+(1 - h_+(t,z))dy^2 +dz^2 
 \label{eqn:metric}
\end{equation}
where $h_+(t,z) = A_+ \cos(\omega_g(t-z))$. We now study the propagation of YM fields described in the previous section [\ref{sec2}] in this background.

We solve the YM Eq. (\ref{eqYM}) in the background of a gravitational wave metric $g_{\mu \nu}$ as in the following:
 \begin{equation}{\label{YMGW}}
     \frac{1}{\sqrt{-g}}\partial_\mu (\sqrt{-g}\ g^{\mu\rho}\ g^{\nu \sigma} F^a_{\rho \sigma}) + g_{\rm{YM}} \epsilon^{abc} \ A^b_\mu \ g^{\mu \rho}\ g^{\nu \sigma} F^c_{\rho \sigma} = 0 
 \end{equation} 
The expression for $F^a_{\mu\nu}$ is given in Eq. (\ref{eqF}).

We then assume the perturbation of initial gauge field $ A^a_\mu = \bar{A}^a_\mu + \Tilde{A}^a_\mu $ where $ \bar{A}^a_\mu $ is the known solution of Yang-Mills equation in the flat background without any source and the $ \Tilde{A}^a_\mu  $ is the perturbation function.
Then we will expand the YM Eq. (\ref{YMGW}) up to linear order in $\Tilde{A}$ and $h_+$. Using $g^{\mu\nu}=\eta^{\mu\nu} + h^{\mu\nu}$, we get (note that this $h^{\mu \nu}=-\eta^{\mu\lambda}\eta^{\nu \rho}h_{\lambda \rho}$ as defined in the metric in Eq. (\ref{eqn:metric}))

\begin{multline}{\label{pert}}
  \eta^{\mu \rho}\eta^{\nu \lambda} \partial_{\mu} \tilde{F}^a_{\rho \lambda}+ g_{\rm{YM}} \epsilon^{abc} ( \eta^{\mu \rho} \eta^{\nu \lambda} \tilde{A}_{\mu}^b \bar{F}^c_{\rho\lambda} + \eta^{\mu \rho} \eta^{\nu \lambda} \bar{A}_{\mu}^b \tilde{F}^c_{\rho\lambda})= - \Big( \eta^{\mu \rho}\partial_{\mu}\left(h^{\nu \lambda} \bar{F}^a_{\rho \lambda}\right)
  +\eta^{\nu \lambda}\partial_{\mu}\left(h^{\mu \rho} \bar{F}^a_{\rho \lambda}\right) \\ + \tilde{\Gamma}^\mu_{\mu\rho} \eta^{\rho\lambda} \eta^{\nu\sigma} \bar{F}^a_{\lambda \sigma} + \tilde{\Gamma}^\nu_{\mu\rho} \eta^{\mu\lambda} \eta^{\rho\sigma} \bar{F}^a_{\lambda \sigma} + 
  g_{\rm{YM}} \epsilon^{abc} \big( h^{\mu \rho} \eta^{\nu \lambda} \bar{A}^b_{\mu} \bar{F}^{c}_{\rho \lambda}+ \eta^{\mu \rho} h^{\nu \lambda} \bar{A}^b_{\mu} \bar{F}^c_{\rho\lambda} \big) \Big)
\end{multline}
where $\bar{F}^a_{\mu\nu}$ is the unperturbed field strength tensor.

Now we will solve the above equation separately for each type of YM wave defined in the previous section [\ref{sec2}]. 
\subsection{Type I waves}{\label{sec3:TypeI}}
We choose the ansatz of type I perturbed waves as discussed in section [\ref{sec2:typeI}]. We take a perturbation in function $U$ as $U=\bar{U} + \tilde{U}$. Even though the function $f^a_\alpha$ is a function of $U$, we choose them to be unperturbed in form or as a function of the total $U$. If we take $f_{\alpha}(\bar{U}+ \tilde{U})=f_{\alpha}(\bar{U}) + \partial f_{\alpha}(U) \tilde{U}$, then only the zeroeth order term contributes to first order in the perturbations as anticipated by the ansatz; 
\begin{equation}{\label{eq:typeI}}
    A^a_\mu = \bar{A}^a_\mu + \tilde{A}^a_\mu =
    \psi_{\alpha} f^a_{\alpha} (\bar{U}) \partial_{\mu} \bar{U} +  \psi_{\alpha} f^a_{\alpha} (U) \partial_{\mu} \tilde{U}
\end{equation}
In the above, we have assumed that the GW modifies the function $U$, as we are require a plane wave induced by the GW as in the zeroeth level ansatz. The $\psi_{\alpha}f^a_{\alpha}$ form in the initial ansatz is motivated from the gauge internal group structure, but GW interaction terms are gauge invariant, and therefore will not change that structure. Therefore, we keep the same form as in the original solution, in the GW induced perturbation.
Using the above ansatz in Eq. (\ref{pert}), we get the following set of in-homogeneous differential equations
\begin{multline}{\label{eq:sec3.1}}
     \p^{\mu}\p^\nu \Tilde{U} \p_\mu \psi_\alpha + \p^\nu \tilde{U} \Box \psi_\alpha -    \p^{\lambda}\p^{\nu} \psi_{\alpha} \p_{\lambda} \tilde{U} - \p^{\nu}\psi_{\alpha} \Box \tilde{U} = \p^\mu \bar{U} \p_\mu (h^{\nu\sigma}\p_\sigma \psi_{\alpha}) + h^{\mu\rho} \p_\rho \bar{U} \p_\mu \p^\nu \psi_\alpha \\ + h^{\nu\sigma} \p_\sigma \psi_\alpha \eta^{\mu\rho} \p_\mu\p_\rho \bar{U} + \p^\nu \psi_\alpha \p_\mu (h^{\mu\rho} \p_\rho \bar{U})  - \eta^{\mu\rho} \p_\mu \p_\rho \psi_\alpha h^{\nu\sigma} \p_\sigma \bar{U} \\ - \p^\nu \bar{U} \p_\mu (h^{\mu\rho} \p_\rho \psi_\alpha) - \eta^{\mu\rho} \p_\rho \psi_\alpha \p_\mu (h^{\nu\sigma}\p_\sigma \bar{U}) - h^{\mu\rho} \p_\rho \psi_\alpha \p_\mu \p^\nu \bar{U}
\end{multline}
with conditions
\begin{align}
     \label{con1:sec3.1}
     \eta^{\mu\nu} \p_{\mu} \tilde{U} \p_\nu \psi_\alpha + h^{\mu\nu} \p_\mu \bar{U} \p_\nu \psi_\alpha & = 0 ,\\
     \label{con2:sec3.1}
    2 \eta^{\mu\nu} \p_{\mu} \tilde{U} \p_\nu \bar{U} + h^{\mu\nu} \p_\mu \bar{U} \p_\nu \bar{U} & = 0 .
\end{align}
These conditions are the same as specified in the previous section for $U,\psi_{\alpha}$ but are now also implemented to first order in the perturbation.

On choosing the different $\psi_\a$, we can find different solutions for $\tilde{U}$.

\subsubsection{Type Ia waves}{\label{sec3:typeIa}}

Using the same assumption as in section [\ref{sec2:typeIa }], we choose $\psi_1 = x$. From the condition (\ref{con1:sec3.1}), one can find that $\p_1 \tilde{U} = 0 $.

Then we were left with only one equation 
\begin{equation}
    \Box\ {\tilde{U}} = \eta^{\mu\rho} \p_\rho \bar{U} \p_\mu h_+ + h_+ \eta^{\mu\rho} \p_\mu \p_\rho \bar{U} - h_+ \p_2 \p_2 \bar{U} 
    \label{eqn:pert1}
\end{equation}

We found that there is no interaction if the waves were propagating in the same direction. If we take the initial waves to be in opposite directions (choosing $\bar{U} = U_0 e^{i\omega_y(t+z)},\omega_y $ is the YM wave frequency), then there is a non-zero in-homogeneous differential equation. However, to implement non-trivial perturbative solutions, we find that we cannot use the condition Eq. (\ref{con2:sec3.1}). To facilitate the solution in the same form as Type Ia, we instead assume $f^a_{\alpha}(U)=c^a_{\alpha}$, or the arbitrary function of $U$ is a constant. The equation of motion, then is of the form in Eq. (\ref{eqn:pert1}) without the condition Eq. (\ref{con2:sec3.1}), on the perturbations. 

The $\tilde{U}$, obtained by solving Eq. (\ref{eqn:pert1})  is
\begin{equation}
    \tilde{U}(t,z) = \frac{1}{2} A_+ U_0 \cos(\omega_g(t-z)) \cos(\omega_y(t+z))
\end{equation}
where $\omega_y$ is the frequency of YM wave.

We further examine the propagation of waves in different orientations.
 
For example, if we choose the $\bar{U}(t,y) = U_0 \cos(\omega_y(t+y))$ i.e. a wave travelling in the -$y$-direction, then the solution for perturbation $\tilde{U}$ is 
\begin{equation}
    \tilde{U}(t,y,z) = -\frac{1}{2} U_0 A_+ \frac{\omega_y}{\omega_g} \Big[\sin(\omega_g(t-z)) \sin(\omega_y(t+y))  \Big] 
\end{equation}

We can see from the condition equations that 
by choosing $\psi_1 = x_1$, we are restricting the propagating direction for YM waves. For example, we can't able to produce an $x$-directed propagating wave since $\p_1 U = 0 $. 


\subsubsection{Type Ib waves}{\label{sec3:typeIb}}
Using the assumption as in section [\ref{sec2:typeIb}], we choose $\Box{\psi}_\alpha = 0 $. Then we get $\psi_\alpha = A_\alpha e^{i \textbf{q}\cdot \textbf{x}}$ and the Eq. (\ref{eq:sec3.1}) gets reduced to 

\begin{equation}
   q^{\nu} \Box \tilde{U} = \eta^{\nu \lambda} h^{\mu \rho} \partial_{\mu}\partial_{\rho} \bar{U} q_{\lambda},
\end{equation}
using some extra conditions $q_\rho \p^\rho h^{\mu \nu} = 0$ and $ q_\lambda h^{\nu \lambda} = 0$ 

From the components of $h_{\mu\nu}$, we find that $q$ vector has two components $q_0 = - q_3 \neq 0$. If we assume that the wave travelling in the -$x$-direction, then $\bar{U}(x,t)= U_0 \cos(\omega_y(x+t))$. In order to compute the perturbative solutions, we choose the same assumption as in type Ia which is by putting a relaxation on the condition (\ref{con2:sec3.1}) by assuming that the arbitrary function ($f^a_\alpha $) is constant. Then, the solution for the perturbation $\tilde{U}$ is found to be
\begin{equation}
    \tilde{U}(x,z,t) = -\frac{1}{2} A_{+} U_{0} \frac{\omega_y}{\omega_g} \Big[ \sin\left(\omega_g(t-z)\right)\sin\left(\omega_y(x+t)\right) \Big ].
\end{equation}

\begin{figure}[ht]
\begin{subfigure}{0.3\textwidth}
\includegraphics[width=5cm, height=5cm]{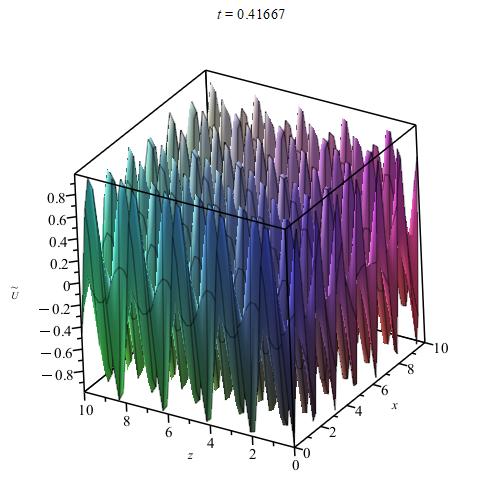} 
\end{subfigure}
\hfill
\begin{subfigure}{0.3\textwidth}
\includegraphics[width=5cm, height=5cm]{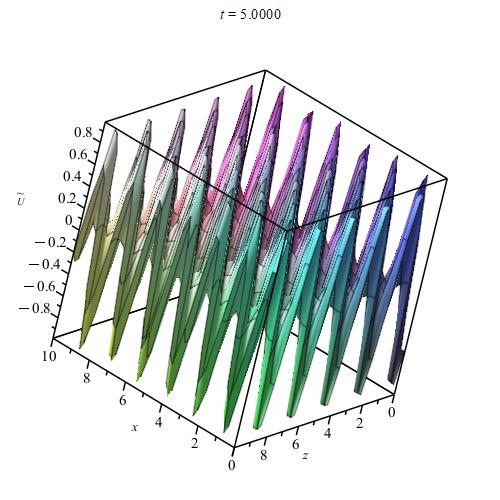}
\end{subfigure}
\hfill
\begin{subfigure}{0.3\textwidth}
\includegraphics[width=5cm, height=5cm]{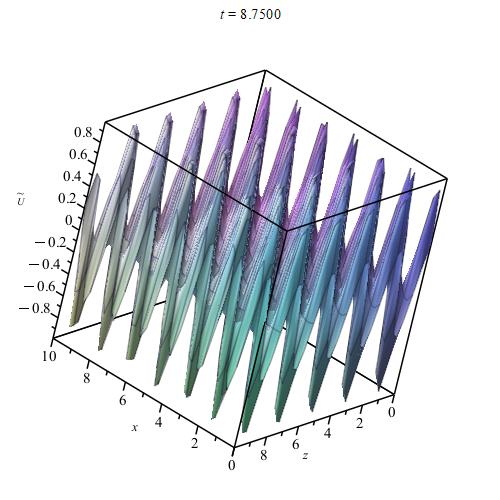}
\end{subfigure}
\hfill
\caption{3d Plot of the YM perturbations due to Gravitational wave with $\omega_g=3$  and $\omega_y=4$.}
\label{fig:1ba}
\end{figure}

\begin{figure}[ht]
\begin{center}
\begin{minipage}[b]{0.45\textwidth}
\centering
\includegraphics[width=6cm,height=5cm]{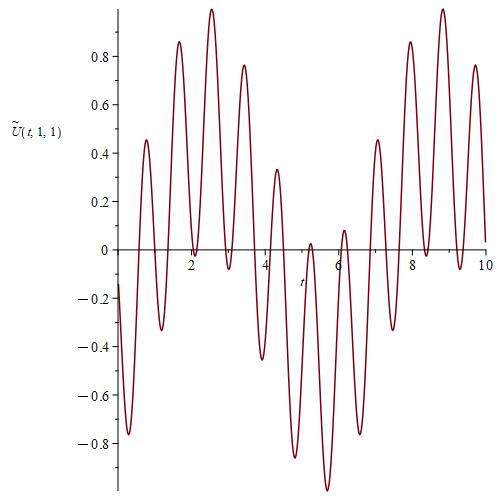}
\end{minipage}
\hfill
\begin{minipage}[b]{0.45\textwidth}
\centering
\includegraphics[width=6cm,height=5cm]{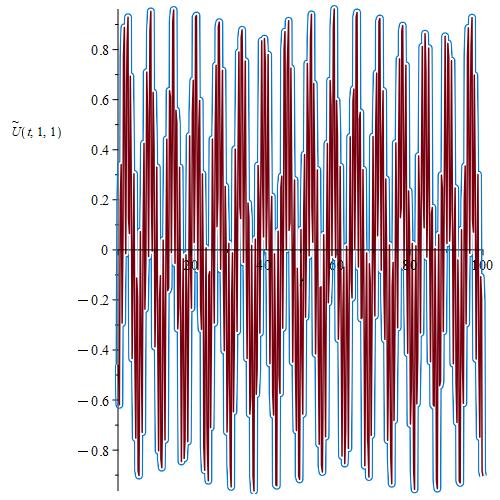}
\end{minipage}
\end{center}
\caption{Plot of the perturbation in time, showing attenuation due to the GWs.}
\label{fig:1b}
\end{figure}

This is a particular solution of the inhomogeneous equation. The plot of the wave shows an overall oscillation in time, superimposed on the oscillations in the two spatial dimensions. A plot in 3 d space shows the pattern, and a plot in pure time isolates the oscillatory profile in time as shown in Fig. (\ref{fig:1ba}\& \ref{fig:1b}). This modulation of the YM wave influences the energy-momentum tensor flow.

As in the Type Ia perturbations, in this case, we find no interaction between the waves travelling in the same direction.

\subsubsection{Coleman waves}{\label{sec3:coleman}}
We will consider the Coleman ansatz for this case. Note that even though in the unperturbed case we can map the Coleman ansatz to the Type I waves, in the calculation of the perturbation, we simply generalize the Coleman form of the zeroth solution.
In order to solve easily using the techniques of \cite{COLEMAN}, we will convert the metric in light-cone coordinates ($x^\pm=t \pm z$) which is 
\begin{equation}
     ds^2 = -dx^- dx^+ + (1 + h_+(x^-))dx^2+(1 - h_+(x^-))dy^2  
 \end{equation}
where $h_+(x^-) = A_+ \cos(\omega_g x^-) $. 
We assume that the Field strength tensors are of the form
 \begin{equation}
 F^a_{\mu \nu}= \bar{F}^a_{\mu \nu} + \tilde{F}^a_{\mu \nu}
 \end{equation}
 and the Gauge fields are of the form $A^a_{\mu}= \bar{A}^a_{\mu} + \tilde{A}^a_{\mu}$, where the $\tilde{F},\tilde{A}$ are the perturbed quantities proportional to the gravitational wave amplitude and $\bar{A}^a_\mu$ is the Coleman solution in lightcone coordinates which can be written as $\bar{A}^a_\mu = \delta^+_\mu\   (f^a(x^+) x + g^a(x^+) y ) $, here $f^a$ and $g^a$ are arbitrary functions.
 
As previously in Coleman's gauge choice, we keep $\tilde{A}_x^a=\tilde{A}_y^a=0$.
 The linearized equations which we can get by substituting $\nu=+,1,2,-$ in Eq. (\ref{pert}) are as follows:
 
 \begin{equation}{ \label{eqn:nu+}}
 \eta^{\mu \rho} \partial_\mu (\tilde{F}_{\rho -}^a)=0
 \end{equation}
 
 \begin{equation}{\label{eqn:nu1}}
 \eta^{\mu \rho} \partial_{\mu}\left(\tilde{F}^a_{\rho 1}\right) + g_{YM} \epsilon^{abc} \eta^{+-} \bar{A}^b_+\tilde{F}^c_{-1} + g_{YM} \epsilon^{abc} \eta^{+-} \tilde{A}^b_-\bar{F}^c_{+1} = -\eta^{-+}\partial_- \left( h^{11} \bar{F}^a_{+1}\right)
 \end{equation}
 
\begin{equation}{\label{eqn:nu2}}
 \eta^{\mu \rho} \partial_{\mu}\left(\tilde{F}^a_{\rho 2}\right) + g_{YM} \epsilon^{abc} \eta^{+-} \bar{A}^b_+\tilde{F}^c_{-2} + g_{YM} \epsilon^{abc} \eta^{+-} \tilde{A}^b_-\bar{F}^c_{+2} = -\eta^{-+}\partial_- \left( h^{22} \bar{F}^a_{+2}\right)
 \end{equation}

\begin{equation} \label{eqn:nu-}
 \eta^{\mu \rho}\partial_{\mu} \left(\tilde{F}^a_{\rho +}\right) + g_{YM} \epsilon^{abc} \eta^{+-} \eta^{-+} \bar{A}_{+}^b\tilde{F}_{-+}^c=0
 \end{equation}

Assume $F^a_{\mu\nu}$ is independent of $x$ and $y$ as in \cite{COLEMAN}. Also, assume the gauge fields are in the same direction in SU(2) internal space. This is consistent with \cite{COLEMAN}. Then the cross terms become zero i.e. the terms involving $\epsilon^{abc}$ are zero. It also means the $a$ index for all the components is the same. Then the above four equations become
 \begin{equation}{\label{nu++}}
 \partial_- (\tilde{F}^a_{+-})=0
 \end{equation}

\begin{equation}{\label{nu11}}
 \partial_-(\tilde{F}^a_{+1}) + \partial_+(\tilde{F}^a_{-1})= - (\partial_- h_{+}(x^-)) f^a(x^+)
\end{equation}

\begin{equation}{\label{nu22}}
 \partial_-(\tilde{F}^a_{+2}) + \partial_+(\tilde{F}^a_{-2})= (\partial_- h_{+}(x^-)) g^a(x^+)
\end{equation}

\begin{equation}{\label{nu--}}
\partial_+(\tilde{F}^a_{-+})=0.
\end{equation}

From equations Eq. (\ref{nu++}) and Eq. (\ref{nu--}), it can be seen that $\tilde{F}^a_{-+} = 0 $ or $\partial_-\tilde{A}^a_+ = \partial_+ \tilde{A}^a_-$. Using this, we can solve for $\tilde{A}^a_+$ and using Eq. (\ref{nu11}) and Eq. (\ref{nu22}). Finally, we can solve for $\tilde{A}^a_-$ using the above condition. 

The final solutions are 

\begin{equation}
 \tilde{A}^a_+  =  \frac12 A_{+}\cos(\omega_g(x^-)) \left[ f^a(x^+) x - g^a(x^+) y\right] 
\end{equation}
\begin{equation}
\tilde{A}^a_- = -\frac12 A_{+}\omega_g \sin(\omega_g(x^-)) \left[ \left(\int f^a(x^+) dx^+\right) x - \left(\int g^a(x^+) dx^+\right) y\right]
\end{equation}
\begin{equation}
    \tilde{A}^a_x = 0
\end{equation}
\begin{equation}
    \tilde{A}^a_y = 0
\end{equation}

In order to understand the nature of solutions, we choose the arbitrary functions $f^a$ and $g^a$ as $f^a(x^+)=g^a(x^+) = \delta^a_1 \sin(\omega_y x^+) $, where $\omega_y$ is the frequency of YM wave. 


\begin{figure}[ht]
\begin{center}
\begin{minipage}[b]{0.45\textwidth}
\centering
\includegraphics[width=6cm,height=5cm]{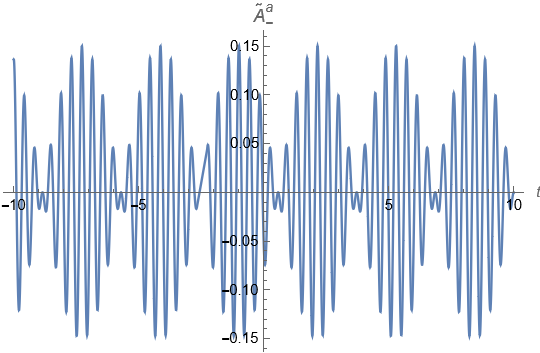}
\end{minipage}
\hfill
\begin{minipage}[b]{0.45\textwidth}
\centering
\includegraphics[width=6cm,height=5cm]{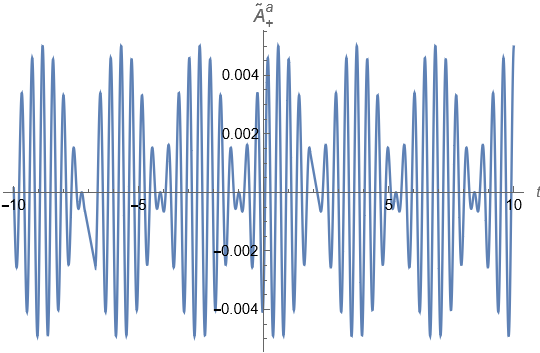}
\end{minipage}
\end{center}
\caption{The plots of $\tilde{A}^a_-(z)\ vs\ z$ and $\tilde{A}^a_+(z)\ vs\ z$. We choose $x=2,\ y=1,\ \omega_g = 15 , \ \omega_y=1  ,\ A_+=0.01\  \textrm{and}\ z= 1 $}
\label{fig:MColeman}
\end{figure}

We can analyze the behaviour of both nonzero solutions by plotting them, $\tilde{A}^a_+$ and $\tilde{A}^a_-$. We choose the frequencies to be $\omega_g = 15  $ and $\omega_y = 1  $ and the plots are given in Fig. (\ref{fig:MColeman}). Also, just to show the functional dependence, we have set the gravitational wave a finite amplitude of $A_+= 0.01$. 

The perturbed non-Abelian electric and magnetic fields can be obtained by computing $E^a_i = F^a_{i0}$ and $B^a_i = \epsilon_{ijk}F^{ajk}$. By changing the light-cone coordinates ($x^+,x^-,x,y$) into Cartesian coordinates ($ t,x,y,z $), we get the following expressions for perturbed electric and magnetic fields (up to first order) 
\begin{equation}
  \tilde{E}^a_x  =  \delta^a_1  \frac{1}{4\omega_y}\big[ 2 A_+ \omega_g \cos(\omega_y(t + z)) \sin(\omega_g(t - z)) + \omega_y (2+A_+ \cos(\omega_g(t - z))) \sin(\omega_y(t +z)\big],
 \end{equation}

\begin{equation}
    \tilde{E}^a_y = \delta^a_1  \frac{1}{4\omega_y}\big[ -2 A_+ \omega_g \cos(\omega_y(t + z)) \sin(\omega_g(t - z)) + \omega_y (2 - A_+ \cos(\omega_g(t - z))) \sin(\omega_y(t +z)\big],
\end{equation}

\begin{equation}
    \tilde{E}^a_z  = - \delta^a_1 \frac{1}{2}A_+ (x - y) \omega_g \sin(\omega_g(t - z)) \sin(\omega_y(t + z)) ,
\end{equation} 

\begin{equation}
     \tilde{B}^a_x  =  \delta^a_1  \frac{1}{2\omega_y}\big[ 2 A_+ \omega_g \cos(\omega_y(t + z)) \sin(\omega_g(t - z)) + \omega_y (2+A_+ \cos(\omega_g(t - z))) \sin(\omega_y(t +z)\big] , 
\end{equation}

\begin{equation}
    \tilde{B}^a_y =  \delta^a_1  \frac{1}{2\omega_y}\big[ 2 A_+ \omega_g \cos(\omega_y(t + z)) \sin(\omega_g(t - z)) + \omega_y (-2+A_+ \cos(\omega_g(t - z))) \sin(\omega_y(t +z)\big],
\end{equation}

\begin{equation}
     \tilde{B}^a_z  = 0.
\end{equation}

We can also calculate the Poynting vector for the perturbed solution from the energy-momentum tensor. The YM energy-momentum tensor is given by 
\begin{equation}
    T_{\mu\nu} =  g^{\rho \sigma}F^a_{\mu\rho} F_{\nu\sigma}^a -\frac{1}{4} g_{\mu\nu} F^a_{\rho \sigma} F^{a \rho \sigma}
\end{equation}

The perturbed non-zero components of the Poynting vector (up to first order) are 
\begin{align}
    S_x = S_y = \frac{1}{4\pi} T_{0x} & = \frac{1}{16\pi}\ A_+\ (x - 
   y)\ \omega_g\ \sin(\omega_g(t - z))\ \sin(\omega_y(t + z))^2 \\
    S_z = \frac{1}{4\pi} T_{0z} & = \frac{1}{8\pi}\ \sin(\omega_y(t + z))^2
\end{align}

\begin{figure}
\centering
\includegraphics[width=\linewidth]{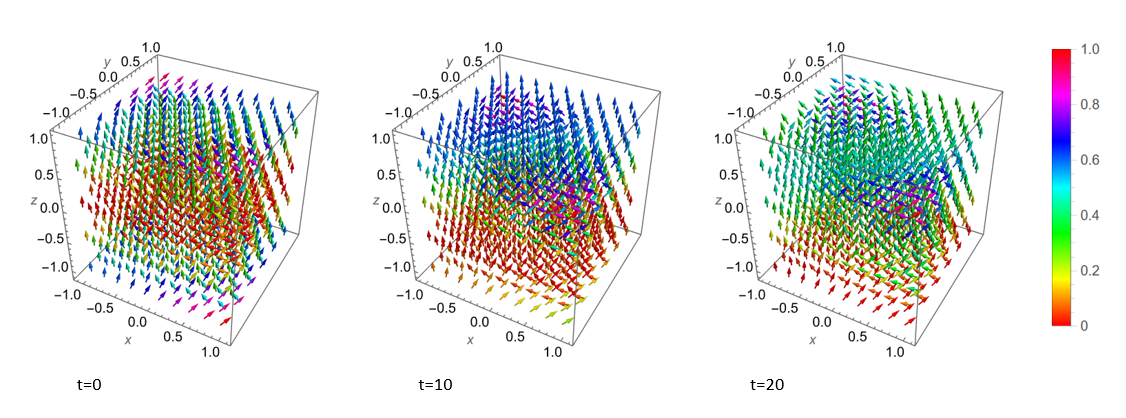}
\caption{Plot of Poynting vector at different times. We choose $\omega_g = 15  $, $\omega_y = 1  $ and $A_+ = 0.1$. }
\label{fig:perToi_coleman}
\end{figure}

We made a field plot (\ref{fig:perToi_coleman}) of the perturbed Poynting vector to show that there is a slight deviation from the original direction.

We further examine the propagation of waves in the same direction and find that there is no interaction. For this, we choose the gauge field of form $\bar{A}^a_x = \bar{A}^a_y = \bar{A}^a_+ =0 $ and $ \bar{A}^a_- = x f^a(x^-)+y g^a(x^-) $, then the source terms in Eq. (\ref{pert}) becomes zero. 

We also examine the interaction of waves propagating in perpendicular directions. We take the YM wave to be propagating in -$x$-direction and the gravitational wave to be in the $z$-direction. In this case, the unperturbed solution is written as  $\bar{A}^a_\mu = \delta^+_\mu\   (f^a(x^+) y + g^a(x^+) z ) $ where $x^+ =t + x $ and $x^- = t -x $. In these new coordinates, the metric becomes
\begin{equation}
     ds^2 = \frac{1}{4} h_+ (dx^+)^2+\frac{1}{4} h_+ (dx^-)^2 - \frac{1}{2}(2+h_+) dx^- dx^+ +(1 - h_+(x^-))dy^2 + dz^2,
\end{equation}
where $h_+ = A_+ \cos(\frac{\om_g}{2}((x^+ + x^-)-2z))$. 

In this case, we assume that the gauge fields point in the same direction in SU(2) space. Because of this, the cross terms in Eq. (\ref{pert}) become zero. Also, using Lorenz gauge ($ \p^\rho \tilde{A}^{a}_{\rho} = 0 $), we can reduce the YM equation (\ref{pert}) into the following four inhomogeneous wave equations which can be solved by assuming some functions for $f^a$ and $g^a$.

\begin{align}
    \Box{\tilde{A}^a_-} & = \frac{1}{2} g^a(x^+) \p_z h_+\\
    \Box{\tilde{A}}^a_y & = 2 f^a(x^+) \p_- h_+ + \p_+(h_+ f^a(x^+))\\
     \Box{\tilde{A}}^a_z & = -g^a(x^+)\p_- h_+ + \p_+(h_+ g^a(x^+))\\
      \Box{\tilde{A}}^a_+ & = - \frac{1}{2} g^a(x^+) \p_z h_+
\end{align}

These are second-order in-homogeneous differential equations. These kinds of equations can be solved using Kirchoff's theorem and Duhamel's principle \cite{Morales,zauderer}. In order to obtain the solutions, we choose the arbitrary functions to be $f^a(x^+) = g^a(x^+) =\delta^a_1 e^{i\omega_y x^+}$. The solutions for the above equations are found to be 

\begin{multline}
    \tilde{A}^a_- = -\tilde{A}^a_+ = \delta^a_1 \frac{ A_+}{4\omega_y} \Big( \cos(\omega_g(t-z)) \sin(\omega_y(t+x)) - \cos(||\omega||t)\cos(\omega_gz)\sin(\omega_yx) \Big.  \\  \Big. + \frac{\sin(||\omega||t)}{||\omega||} \left( \omega_y \cos(\omega_yx)\cos(\omega_gz) - \omega_g \sin(\omega_yx)\sin(\omega_gz) \right)  
    \Big)
\end{multline}

\begin{multline}
    \tilde{A}^a_y = \delta^a_1  \frac{A_+}{4\omega_y\omega_g} \Bigg( \frac{\sin(||\omega||t)}{||\omega||} \Big[ 5\omega_g\omega_y\cos(\omega_yx)\cos(\omega_gz) + (3\omega_g^2+2\omega_y^2)\sin(\omega_yx)\sin(\omega_gz) \Big] \Bigg.  \\  \Bigg.
    +\cos(||\omega||t) \Big[ 3\omega_g\sin(\omega_yx)\cos(\omega_gz) - 2 \omega_y \cos(\omega_yx)\sin(\omega_gz) \Big] - 2\omega_y\sin(\omega_g(t-z)) \cos(\omega_y(t+x)) \Bigg. \\ \Bigg. 
    - 3\omega_g \cos(\omega_g(t-z))\sin(\omega_y(t+x)) \Bigg)
\end{multline}

\begin{multline}
    \tilde{A}^a_z = \delta^a_1 \frac{ A_+}{2\omega_g} \Bigg( \frac{\sin(||\omega||t)}{||\omega||} \Big[ \omega_g\cos(\omega_gz)\cos(\omega_yx) + \omega_y \sin(\omega_gz)\sin(\omega_yx) \Big] \Bigg. \\  \Bigg.  -\sin(\omega_g(t-z)) \cos(\omega_y(t+x))- \cos(||\omega||t) \cos(\omega_yx) \sin(\omega_gz) \Bigg)
\end{multline}
where $||\omega|| = \sqrt{\omega_g^2+\omega_y^2}$.

In the above, we have used Green's function for solutions to mass-less inhomogeneous wave equations as described in \cite{Morales}. These solutions, apart from the particular solution to the inhomogeneous equations, also have contributions from modes with frequency $\sqrt{\omega_g^2+\omega_y^2}$ as in \cite{Akash, Morales}.  If we require the boundary condition that the perturbations are zero at the initial time, i.e. t=0, then the Green's function gives the total perturbation to comprise the particular and the general solution. The frequency $\sqrt{\omega_g^2 + \omega_y^2}$ appears due to the solution of the homogeneous differential equation as observed in \cite{Akash}. In an actual experiment, it is a natural boundary condition that the perturbation is 0, before the time of the arrival of the GW signal. Once the GW reaches the laser beam, the perturbation is generated at t=0.  It shall be interesting to detect the $\sqrt{\omega_g^2+\omega_y^2}$ in an actual observation of the matter wave responding to a GW.
In the solutions for Type I a and I b waves, we did not use the boundary conditions of \cite{Morales, Akash}, but obtained the particular solutions to the in-homogeneous differential equations. This is because based on the nature of the ansatz the boundary condition could not be imposed. This means that the Coleman ansatz is the correct one to describe the YM wave and GW interaction in a detector.

\subsection{Type II waves}{\label{sec3:typeII}}
As we can see in the previous section, the perturbations don't involve the YM coupling constant ($g_{\rm{YM}}$). In order to include the YM coupling constant in solutions, we have to take a different ansatz. First, we will take the metric in usual Cartesian coordinates as in Eq. (\ref{eqn:metric}). We assume the gauge fields of the form 
\begin{equation}{\label{eq:ans2}}
    A^a_\mu = \bar{A}^a_\mu + \tilde{A}^a_\mu
\end{equation}
where $ \bar{A}^a_\mu $ is the Type II YM wave solution (\ref{ansatz2}) in Minkowski background and $\tilde{A}^a_\mu$ is the perturbation function.

Now we will restrict the propagation of unperturbed YM solution (\ref{ansatz2}) by choosing the $p_\mu$ vector to be $p_\mu = (p_0,0,0,p_3) $ such that $p_\mu$ is time like. By choosing the $p_0$ and $p_3$, we can make the wave propagate in  $z$ or $-z$ direction.

From the inspiration from the unperturbed solution, we can assume the perturbation function to be  
\begin{equation}{\label{eq:pertansatz2}}
    \tilde{A}^a_\mu= \delta^a_1\ \tilde{\Phi}\ k_\mu + \delta^a_3\ \tilde{\Psi}\ l_\mu
\end{equation}  
where $\tilde{\Phi}$ and $\tilde{\Psi}$ are functions of $x^\mu$.
Here we have taken the perturbation ansatz to have the same form as the unperturbed YM solution motivated from the fact that the GW propagates in the +z, or -z direction. The GW polarizations interact with the YM polarizations in the $l_{\mu}$ and $k_{\mu}$ directions only. If we take a GW propagating in a direction perpendicular to the YM wave, let us say in the x-direction, the equations are much complicated and we cannot use the above ansatz.  We have to introduce new functions apart from the $\tilde{\Phi}, \tilde{\Psi}$. These require an entire independent analysis and is outside the scope of this current paper.
By substituting the metric (\ref{eqn:metric}), unperturbed solution (\ref{ansatz2}) and perturbation function (\ref{eq:pertansatz2}) in Eq. (\ref{pert}), we will get the equations for $\tilde{\Phi}$ and $\tilde{\Psi}$. The nature of solutions also depends on the choice of $k_\mu$ and $l_\mu$ vectors. We choose the vectors such that they satisfy $k_\mu \p^\mu \Phi = l_\mu \p^\mu \Phi = k_\mu \p^\mu \Psi = l_\mu \p^\mu \Psi = 0 $. For suppose, we can choose $k_\mu = (0,0,1,0)$ and $l_\mu = (0,1,0,0)$. With this choice, we also get the functional dependence of $\tilde{\Phi}$ and $\tilde{\Psi}$ from the conditions mentioned above (up to the first order). Finally, we will get the following two equations
\begin{align}
     \label{eq:typeIIGW1}
     \Box \tilde{\Phi} - 2 g^2_{\rm{YM}} l^2\; \bar{\Psi}\; \bar{\Phi}\; \tilde{\Psi} - g^2_{\rm{YM}} l^2 \; \bar{\Psi}^2 \; \tilde{\Phi} & = - \p_\mu h_+(t,z) \p^\mu \bar{\Phi} - h_+(t,z) (g_{\rm{YM}}^2 l^2 \; \bar{\Phi}\; \bar{\Psi}^2) \\ 
    \label{eq:typeIIGW2} 
    \Box \tilde{\Psi} - 2 g^2_{\rm{YM}} k^2 \bar{\Psi}\; \bar{\Phi}\; \tilde{\Phi} - g^2_{\rm{YM}} k^2\; \bar{\Phi}^2\; \tilde{\Psi} & =   \p_\mu h_+(t,z) \p^\mu \bar{\Psi} + h_+(t,z) (g_{\rm{YM}}^2 k^2 \; \bar{\Psi}\; \bar{\Phi}^2) 
\end{align}
where $\bar{\Phi}$ and $\bar{\Psi}$ are the unperturbed YM wave solutions (\ref{eq:typeIIsol}).

\begin{figure}[ht]

\begin{subfigure}{0.45\textwidth}
\includegraphics[width=0.8\linewidth,height=4cm]{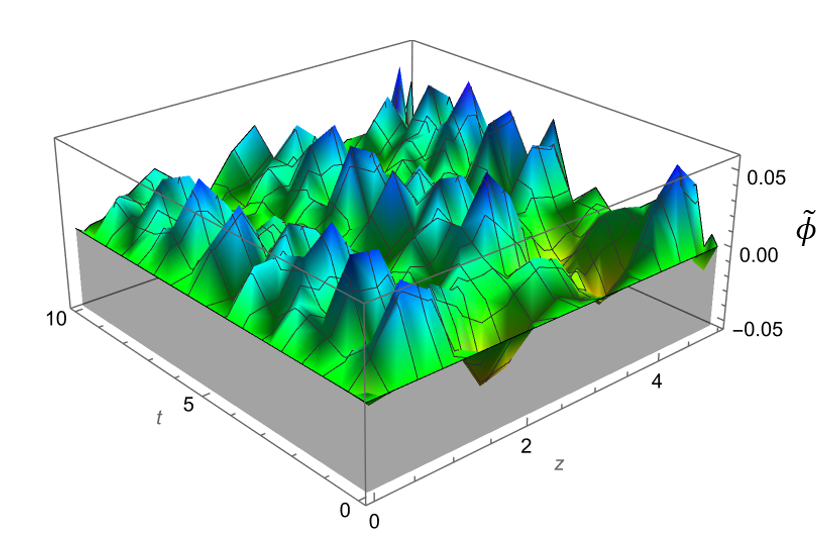}
\caption{$\tilde{\Phi}$}
\end{subfigure}
\begin{subfigure}{0.45\textwidth}
\includegraphics[width=0.8\linewidth, height=4cm]{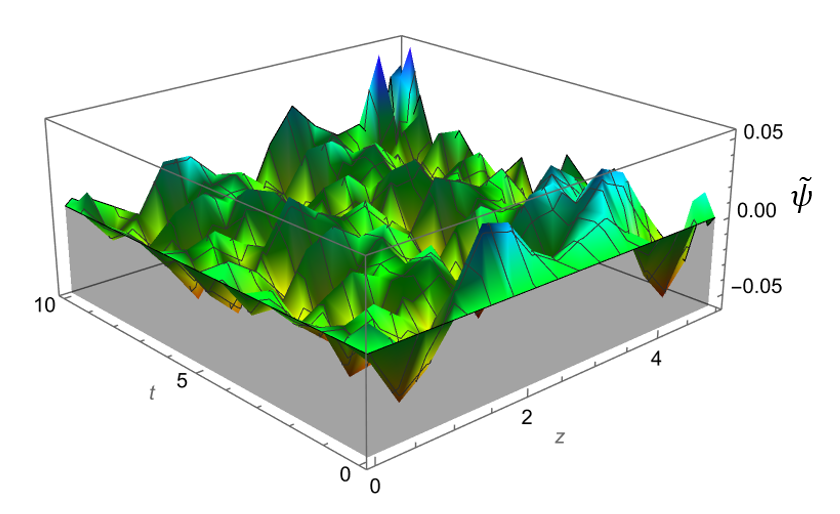}
\caption{$\tilde{\Psi}$}
\end{subfigure}
\caption{Figures showing $\tilde{\Phi}$ and $\tilde{\Psi}$ as a function of t and z. We choose $l_1 = 1$, $k_2 =1$, $A_+ = 0.01$, $g = 0.5$ and $\omega_g=20$. }
\label{fig:MTypeII}
\end{figure}

\begin{figure}[ht]

\begin{subfigure}{0.45\textwidth}
\includegraphics[width=0.8\linewidth,height=4cm]{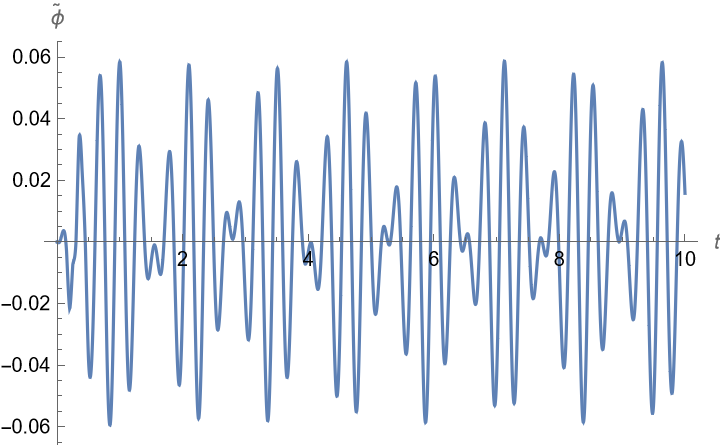}
\caption{$\tilde{\Phi}(1,t)$}
\end{subfigure}
\begin{subfigure}{0.45\textwidth}
\includegraphics[width=0.8\linewidth, height=4cm]{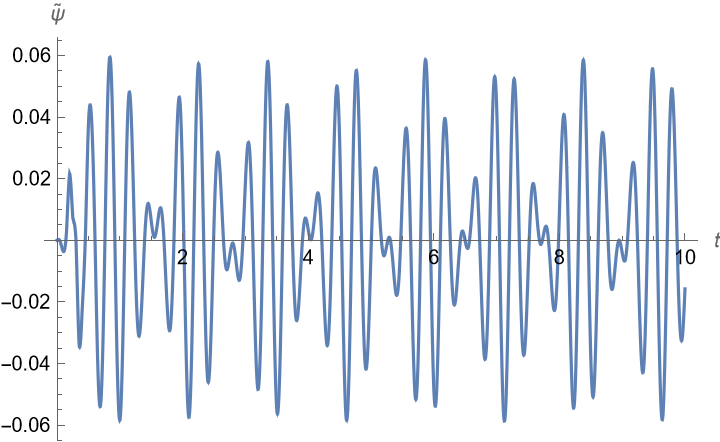}
\caption{$\tilde{\Psi}(1,t)$}
\end{subfigure}
\caption{Figures showing $\tilde{\Phi}$ and $\tilde{\Psi}$ as a function of t. We choose $z=1$, $l_1 = 1$, $k_2 =1$, $A_+ = 0.01$, $g = 0.5$ and $\omega_g=20$. }
\label{fig:perphipsi}
\end{figure}

\begin{figure}
    \centering
    \includegraphics{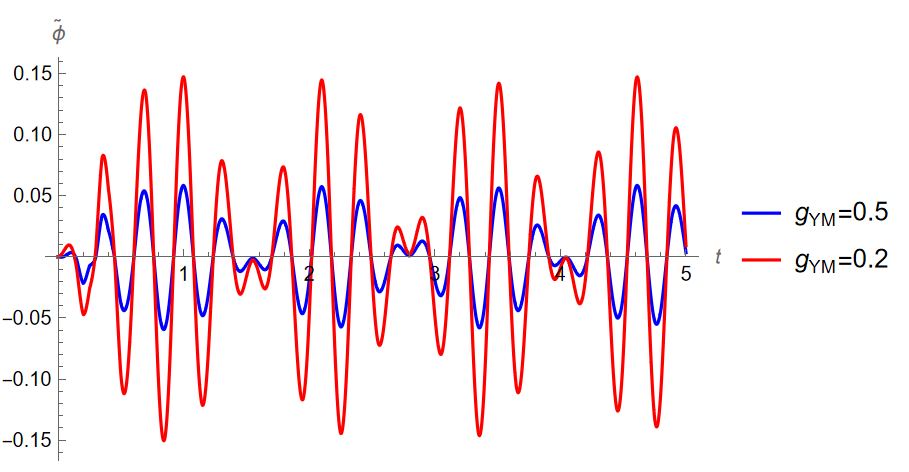}
    \caption{Figure showing $\tilde{\Phi}(1,t)$ for two different values of $g_{\rm{YM}}$.}
    \label{fig:MtypeIIphig}
\end{figure}

We solved the equations Eq. (\ref{eq:typeIIGW1} \& \ref{eq:typeIIGW2}) numerically and plots are shown in Fig. (\ref{fig:MTypeII}). We also show the plots of perturbations of $\Phi$ and $\Psi$ as a function of time ($t$) in Fig. (\ref{fig:perphipsi}). For this, we choose the YM wave propagating in the $-z$ direction by choosing $p_0=3$ and $p_3=1$. We also investigated the dependence of the coupling constant ($g_{\rm{YM}}$) on the solutions. We find that decreasing the coupling constant increases the magnitude of the perturbed function, but the functional behaviour remains the same. This is observed in Fig. (\ref{fig:MtypeIIphig}). This can be understood in the following way: Since the unperturbed functions ($\bar{\Phi}\; \&\; \bar{\Psi}$) are inversely proportional to the coupling constant, it increases the magnitude of unperturbed functions. In the above differential equations, the terms involving coupling constant are always accompanied by the product of two unperturbed functions which renders its influence on functional behaviour.

In order to understand the oscillation frequency of the perturbed functions ($\tilde{\Phi}(1,t) \; \& \; \tilde{\Psi}(1,t)$), we did the spectrum analysis for the perturbed functions in Minkowski spacetime.

\begin{figure}[ht]
\begin{subfigure}{0.45\textwidth}
\includegraphics[width=0.8\linewidth,height=4cm]{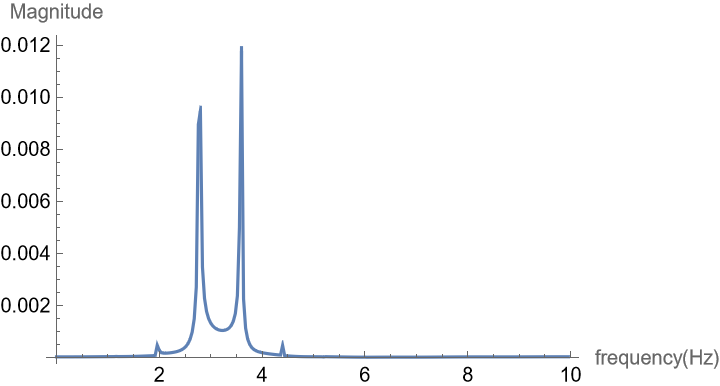}
\caption{Power Spectrum}
\end{subfigure}
\begin{subfigure}{0.45\textwidth}
\includegraphics[width=0.8\linewidth, height=4cm]{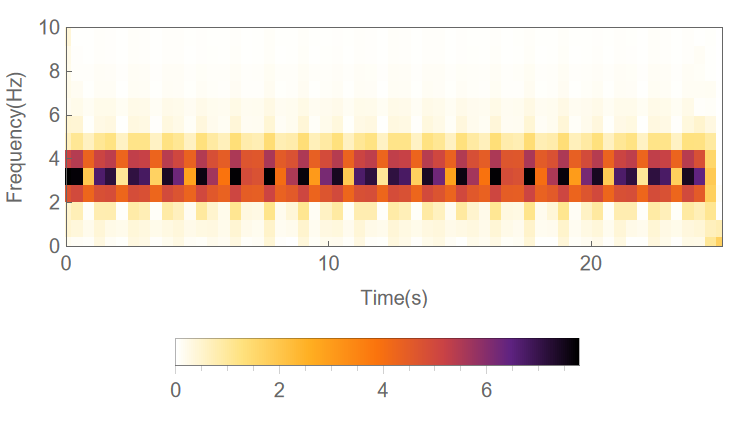}
\caption{Spectrogram}
\end{subfigure}
\caption{Figures showing Power spectrum and spectrogram of $\tilde{\Phi}(1,t)$  }
\label{fig:app}
\end{figure}

\begin{figure}[ht]
    \centering
    \includegraphics[width=0.6\linewidth, height=6cm]{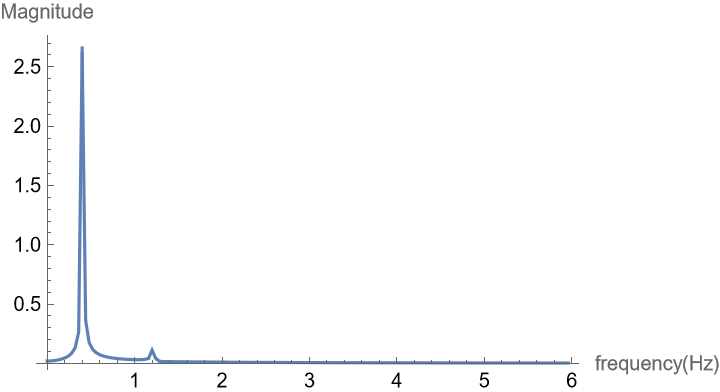}
    \caption{Power Spectrum of cn(3t+1,1/2)}
    \label{fig:cnpower}
\end{figure}

We find that they follow a beats pattern with the inner frequency being the sum of GW and of YM wave frequencies and the outer frequency being the difference between GW and YM wave frequencies. This can be seen in Fig. (\ref{fig:perphipsi}). The power spectrum and spectrogram of $\tilde{\Phi}(1,t)$ are shown in Fig. (\ref{fig:app}). One can see that there are multiple peaks present in the power spectrum. This can be understood in the following way: Since the Cosine Jacobi elliptic functions are not similar to simple sine or cosine functions, they are combinations of multiple sine or cosine functions (Fig. \ref{fig:cnpower}). These frequencies combine with the single GW frequency to give the resultant power spectrum (Fig. \ref{fig:app}).

Evaluating perturbed Poynting vector (up to first order) gives

\begin{multline}
    S_i = \frac{1}{4\pi} T_{0i}  = \delta^3_i \ \frac{1}{4\pi} \big[k_2^2 \partial_0 \bar{\Phi} \ \partial_3 \bar{\Phi} (1+ A_+ \cos(\omega_g(t-z)))+ l_1^2 \partial_0 \bar{\Psi}\ \partial_3 \bar{\Psi} \\
    (1 - A_+ \cos(\omega_g(t-z))) + k_2^2  (\partial_0 \bar{\Phi}\ \partial_3 \tilde{\Phi}+\partial_0 \tilde{\Phi}\ \partial_3 \bar{\Phi})+ l_1^2 (\partial_0 \bar{\Psi}\ \partial_3 \tilde{\Psi}+\partial_0 \tilde{\Psi}\ \partial_3 \bar{\Psi}) \big]
\end{multline}

and the energy density is of form
\begin{multline}
    T_{00} = (1+A_+ \cos(\omega_g(t-z))) (\partial_0 \bar{\Phi} k_2 + \partial_0 \tilde{\Phi} k_2)^2 +(1 - A_+ \cos(\omega_g(t-z))) (\partial_0 \bar{\Psi} l_1 + \partial_0 \tilde{\Psi} l_1)^2 -\\
    \frac{1}{2} (1+A_+ \cos(\omega_g(t-z))) [(\partial_0 \bar{\Phi} k_2 + \partial_0 \tilde{\Phi} k_2)^2-(\partial_3 \bar{\Phi} k_2 + \partial_3 \tilde{\Phi} k_2)^2] - g_{\rm{YM}}^2\ (1+A_+ \cos(\omega_g(t-z)))\\
    (1 - A_+ \cos(\omega_g(t-z)))(\bar{\Phi} k_2 +\tilde{\Phi} k_2 )^2(\bar{\Psi} l_1 + \tilde{\Psi} l_1)^2 + (1 - A_+ \cos(\omega_g(t-z))) [(\partial_0 \bar{\Psi} l_1 + \partial_0 \tilde{\Psi} l_1)^2-(\partial_3 \bar{\Psi} l_1 + \partial_3 \tilde{\Psi} l_1)^2].
\end{multline}

The solutions can be interpreted as non-Abelian plane waves propagating in a fixed z direction since  the energy density is bounded throughout space-time and the direction of the Poynting vector is constant.

Similarly, we find the non-Abelian electric and magnetic fields for the perturbed solution as
\begin{equation}
    \tilde{E}^a_i = -\big[\delta^a_1 \delta^2_i (\partial_0 \tilde{\Phi} k_2)+ \delta^a_3 \delta^1_i (\partial_0 \tilde{\Psi} l_1)   \big]
\end{equation}

\begin{multline}
      \tilde{B}^a_i  = \delta^a_1 \delta^1_i \big[-\partial_3 \tilde{\Phi}\ k_2 - A_+\ \cos(\omega_g(t-z)) \partial_3 \bar{\Phi}\ k_2\big] +\\
          \delta^a_2\delta^3_i \big[g_{\rm{YM}}\ (\phi k_2\ \tilde{\Psi} l_1 + \tilde{\Phi} k_2 \bar{\Psi} l_1 - A_+^2 \cos(\omega_g(t-z))^2 \bar{\Phi} \bar{\Psi} k_2 l_1 ) \big] + \delta^a_3\delta^2_i \big[-\partial_3 \tilde{\Psi} l_1\big]
  \end{multline}
which are real throughout space-time. We find that the electric fields are linear in YM potentials and also transverse to the propagating direction. But the magnetic field which is nonlinear in YM potentials is not perpendicular to the direction of propagation. It can be seen as 
\begin{equation}
    \tilde{B}^a_i \delta^3_i = \tilde{B}^a_3 = g_{\rm{YM}}\ \delta^a_2 \big[ \bar{\Phi} k_2\ \tilde{\Psi} l_1 + \tilde{\Phi} k_2 \bar{\Psi} l_1 - A_+^2 \cos(\omega_g(t-z))^2 \bar{\Phi} \bar{\Psi} k_2 l_1  \big]
\end{equation}

This is not surprising, as the YM fields are not required to be transverse waves from the construction.
\subsection{Summary}
In Section 3, we have solved the propagation of YM fields in the presence of GW in flat space-time. Depending on the nature of the YM progressive wave, labeled as Type I, Coleman, and Type II,  we obtain different perturbations induced by the GW.  The Type I and Coleman waves have the commutator term in the YM field equations as zero, due to the nature of the ansatz. This, therefore, is maintained in the GW background and the perturbation shows patterns similar to `beats'.  The Poynting vector shows changes in a direction similar to EM wave perturbations obtained in \cite{Akash}. We also find that there is no interaction between GW and YM waves when propagating in the same direction for the Type I waves similar to EM wave analysis \cite{Akash}. This isn't surprising that without the commutator terms, the YM equations will be similar to Maxwell equations. The Type II waves on the other hand have non-zero commutators, and due to the linearized approximation in the perturbed equation of motion, the perturbation equation is linear and does not show non-linear interactions. However, the solutions could not be obtained analytically, and have been shown as numerical graphs. The spectral analysis of the numerical solutions shows a similar pattern as the Type I waves, there is a dominance of the frequency of the gravitational wave and the periodicity of the Jacobi-functions of the unperturbed YM fields. {\it The Type II perturbations though have a clear dependence on the SU(2) coupling constant $g_{\rm YM}$. The amplitude of the waves increases as the $g_{\rm YM}$ decreases.} Also, we find that there is an interaction between YM fields and GW even when propagating in the same direction, but only for Type II waves. It is because of the non-linear nature of YM equations. 
 In the next section we investigate how the YM and the GW interaction happens in the cosmological background.
\section{Interaction of Yang-Mills waves with Gravitational waves in Cosmological background }

Study of Gravitational waves in the flat background is a valid approximation if the gravitational waves originated in the late universe. But if we want to study the gravitational waves that originated in the early universe, we need to move away from the flat background approximation. It is well known that there are some processes during the early universe that can generate gravitational waves (PGWs). These PGWs are stochastic in nature and form a background of gravitational waves today, though not yet detected. Even though they form a background, they haven't been detected yet due to their very small amplitude and low frequencies. In this paper, we study the interaction of YM waves with gravitational waves in de-Sitter background. Though de-Sitter space is relevant for inflationary cosmology only, we study it as an example of cosmological background. The gravitational wave is also easy to work with in this background written in conformal time. There are other cosmological backgrounds, in particular radiation dominated and matter dominated ones, and we hope to work on this in the future. For this, we consider the FLRW metric in conformal coordinates as 
\begin{equation}
    ds^2 = a^2(\tau) (-d\tau^2 + dx^2 + dy^2 + dz^2) 
\end{equation}
where x,y,z are Cartesian coordinates and $a(\tau)$ is the scale factor in conformal time coordinate ($\tau$). The conformal time ($\t$) is related to co-moving time ($t$) by relation $d\tau = {dt}/{a(t)} $. We have solved it in the de-Sitter background as it represents the metric during inflation.

We find that the YM equation (\ref{eqYM}) is independent of the scale factor in the conformal coordinates. This can be seen from the equation.

\begin{equation}{\label{eqcos}}
 \frac{1}{\sqrt{-g}} \partial_{\mu} \left(\sqrt{-g} g^{\mu \lambda} g^{\nu \rho} F^a_{\lambda \rho} \right) + g_{\rm{YM}} \epsilon^{abc} A_{\mu}^b g^{\mu \rho} g^{\nu \lambda} F^c_{\lambda \rho} =0
 \end{equation}

If we consider a homogeneous and an isotropic cosmological metric like  $g^{\mu \nu}= \frac{1}{a^2(\tau)}\eta^{\mu \nu}$, then one finds that the above equation reduces to that of same as in flat geometry. This shows that the YM equations are conformally invariant.

Since the YM equations are conformally invariant, their solutions on the cosmological background  have the same form as in the Minkowski background with  the coordinate time ($t$) replaced with the conformal time ($\tau$).

In addition, we consider a gravitational wave with only one polarisation $h_+$ propagating at frequency $\omega_g$ with an amplitude $A_+$ in $z$-direction. Then the metric becomes

\begin{equation}
     ds^2 = a^2(\tau)\big(-d\tau^2 + (1 + h_+(\tau,z))dx^2+(1 - h_+(\tau,z))dy^2 +dz^2 \big ).
\end{equation}
The explicit expressions for $h_+$ waveforms are different in different cosmological backgrounds. In this paper, we consider the gravitational waves in the de-Sitter background.
The exact form is given by 
\begin{equation}{\label{eq:GWexp}}
    h_+(\tau,z) = A_+ (\omega_g \tau - i) e^{-i\omega_g(\tau - z)}.
\end{equation}
For the details of the solution for gravitational waves, see \cite{GW3Allen,GW2Allen,GW1}. By inserting the Hubble parameter, the constant $A_+$ will be proportional to $H_{dS}/\omega_g $. Note that if we work with these solutions, we cannot use a simple $H_{dS}\rightarrow 0$ limit to find the Minkowski space-time. This is because $\tau\propto -1/H_{dS}$, and therefore the $\tau\rightarrow \infty$. The conformal time exists only for $H_{dS}\neq 0$ limit. 

We use the same procedure as described in the section [\ref{sec3}] for the study of the perturbations induced by GW in de-Sitter space.

\subsection{Type I waves}

We will follow the same procedure as described in section [\ref{sec3:TypeI}]. We will take the ansatz of form Eq. (\ref{eq:typeI}). We will get the same differential equations (\ref{eq:sec3.1}) that we got in Minkowski background with only difference being that co-moving time ($t$) is replaced with conformal time ($\t$). By choosing different $\psi$, we get different solutions.

\subsubsection{Type Ia waves}
By choosing $\psi_1= x,\psi_{2,3,4}=0$, the YM equation reduces to a single second-order inhomogeneous differential equation.

\begin{equation}{\label{eq:dIa}}
    \Box{\tilde{U}} = \eta^{\mu\rho} \p_\rho \bar{U} \p_\mu h_+ + h_+ \eta^{\mu\rho} \p_\mu \p_\rho \bar{U} - h_+ \p_2 \p_2 \bar{U} 
\end{equation} 
where $h_+$ is given by Eq. (\ref{eq:GWexp}), however, the conditions derived for the perturbations in Eq. (\ref{con1:sec3.1}) and Eq. (\ref{con2:sec3.1}) remain the same. Since the equation for $h_+$ is not simply sinusoidal, the source terms of Eq. (\ref{eq:dIa}) will not be zero even if we choose the YM wave is propagating in the $z$-direction i.e. ($\bar{U} = U_0 \cos(\omega_y(\t-z))$). It means we have an interaction between the YM wave and GW even when they are propagating in the same direction. Now, let's choose the YM wave travelling in anti-parallel direction i.e. YM wave travelling in -$z$-direction ($\bar{U} = U_0 \cos(\omega_y(\t+z))$). To get the perturbative solutions, we choose the same assumption as we took in Type Ia waves in Minkowski spacetime. Then, we get the solution for $\tilde{U}$ as 

\begin{multline}
    \tilde{U}(\t,z) = -\frac{1}{4\omega_y} A_+ U_0 \Bigg( (\omega_g+2\omega_y) \sin((\omega_y-\omega_g)z+(\omega_g+\omega_y)\t) +(\omega_g-2\omega_y) 
    \Bigg. \\ \Bigg. 
    \sin((\omega_y-\omega_g)\t+(\omega_y+\omega_g)z)-2\t \omega_y\omega_g \Big[ \cos((\omega_y+\omega_g)z+(\omega_y-\omega_g)\t) 
    \Bigg. \Big. \\ \Big. \Bigg. 
     +\cos((\omega_y+\omega_g)\t+(\omega_y-\omega_g)z)  \Big] \Bigg)
\end{multline}
where $\om_y$ is YM wave frequency and $U_0$ is the amplitude of YM wave.

\subsubsection{TypeIb waves}
By choosing $\Box{\psi}_\alpha = 0 $, the YM equation reduces to 
\begin{equation}
   q^{\nu} \Box \tilde{U} = \eta^{\nu \lambda} h^{\mu \rho} \partial_{\mu}\partial_{\rho} U q_{\lambda}
\end{equation}
Here  we choose the YM wave travelling in the $x$-direction, i.e. $\bar{U}(x,\t)= U_0 \cos(\omega_y(x-\t))$. In order to compute the non-trivial perturbative solutions, we relax the condition on $\tilde{U}$ as in Eq.(\ref{con2:sec3.1}) assuming the arbitrary function ($f^a_\alpha $) is constant.Then, we find the solution for $\tilde{U}$ to be  
\begin{equation}
    \tilde{U}(\tau,x,z) = \frac{ A_+ U_0 }{2\omega_g}  
    e^{-i \omega_g(\t-z)} \Big( i \omega_g\cos(\omega_y(\t-x))+(2+i\t \omega_g)\omega_y \sin(\omega_y(\t-x)) \Big) 
\end{equation}

Similarly, in this case, the interaction between YM waves and GWs propagating in the same direction is non zero. 

\subsubsection{Coleman waves}
We will follow the same procedure as described in the section [\ref{sec3:coleman}]. Consider the metric in light cone coordinates ($x^\pm = \tau \pm z $) as
\begin{equation}
     ds^2 = a^2(\tau)\Big( -dx^- dx^+ + (1 + h_+(x^+,x^-)) dx^2+(1 - h_+(x^+,x^-)) dy^2  \Big)
 \end{equation}
where $h_+(x^+,x^-)$ is given by Eq. (\ref{eq:GWexp}) in these coordinates. We will follow the same approach as considering $A^a_\mu = \bar{A}^a_\mu + \tilde{A}^a_\mu$ and $F^a_{\mu \nu}= \bar{F}^a_{\mu \nu} + \tilde{F}^a_{\mu \nu}$ where $\bar{A}$ and  $\bar{F}^a_{\mu \nu} $ are the Coleman solution in cosmological background and $\tilde{A}^a_\mu$ and $ \tilde{F}^a_{\mu \nu} $ are the perturbed quantities. We can choose the gauge $\tilde{A}^a_1 = \tilde{A}^a_2 = 0 $. Also, use the same assumptions in section [\ref{sec3:coleman}] which are (i) $F^a_{\mu\nu}$ is independent of $x$ and $y$ and (ii) The gauge field components point in the same direction in internal SU(2) space. With these, have to solve equations Eq. (\ref{nu++}) - Eq. (\ref{nu--}) and a condition: $\partial_-\tilde{A}^a_+ = \partial_+ \tilde{A}^a_-$ as previously due to conformal invariance. From Eq. (\ref{nu11}) and Eq. (\ref{nu22}), we get $\tilde{A}^a_+$ as

\begin{equation}
 \tilde{A}^a_+  =  \frac12 A_{+} \Big[\frac{\omega_g}{2}(x^++x^-) - i \Big] \Big[ f^a(x^+) x - g^a(x^+) y\Big] e^{-i\omega_gx^-}
\end{equation}
From condition, we get $\tilde{A}^a_-$ as 

\begin{equation}
\tilde{A}^a_- = -\frac14 A_+ \omega_g \left[ \left(\int [1+i \omega_g (x^+ + x^-)]f^a(x^+) dx^+\right) x - \left(\int [1+i \omega_g (x^+ + x^-)] g^a(x^+) dx^+\right) y\right] e^{-i \omega_g x^-}
\end{equation}

\begin{figure}[ht]
\begin{center}
\begin{minipage}[b]{0.45\textwidth}
\centering
\includegraphics[width=6cm,height=5cm]{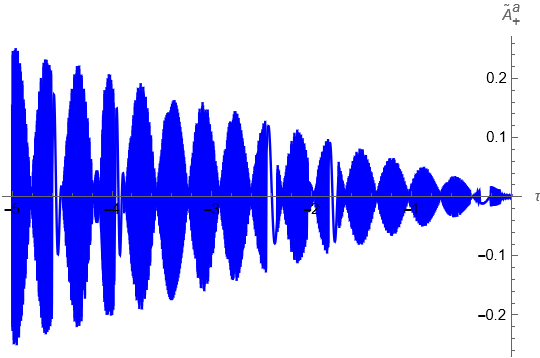}
\end{minipage}
\hfill
\begin{minipage}[b]{0.45\textwidth}
\centering
\includegraphics[width=6cm,height=5cm]{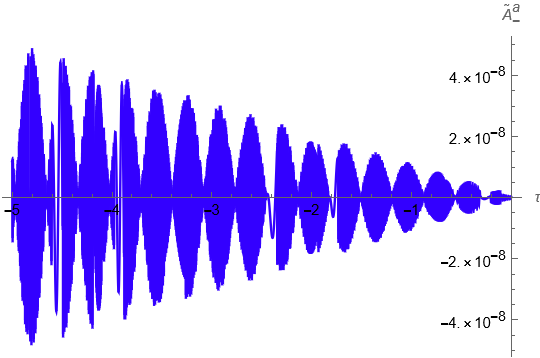}
\end{minipage}
\end{center}
\caption{The perturbed gauge fields as a function of conformal time. We choose $x=2,\ y=1,\ \omega_g = 10 , \ \omega_y=10^8 ,\ A_+=0.01\  \textrm{and}\ z = 1 $.}
\label{fig:Dcoleman}
\end{figure}

We will study the behaviour of nonzero solutions as a function of conformal time. We choose the arbitrary functions $f^a$ and $g^a$ as $f^a(x^+)=g^a(x^+) = \delta^a_1 \sin(\omega_y x^+) $, where $\omega_y$ is the frequency of YM wave. We choose the frequencies to be $\omega_g = 10 $ and $\omega_y = 10^8$ and the plots are given in Fig. (\ref{fig:Dcoleman}). One can see that the magnitude of the solutions decreases with increasing time. This decreasing pattern is identified with $1/\t$ behaviour. 

\subsection{Type II waves}
We will follow the same procedure as described in section [\ref{sec3:typeII}]. We start by assuming the gauge fields of the form Eq. (\ref{eq:ans2}) with $\bar{A}^a_\mu$ be the solution of Type II YM wave propagating in the -$z$-direction (as discussed in section [\ref{sec3:typeII}]) in the cosmological background and the perturbation function is given by Eq. (\ref{eq:pertansatz2}). The nature of solutions also depends on the choice of $k_\mu$ and $l_\mu$ vectors. By choosing the vectors to be $k_\mu = (0,0,1,0)$ and $l_\mu = (0,1,0,0)$, we will finally get to the same form of equations Eq. (\ref{eq:typeIIGW1}) and  Eq. (\ref{eq:typeIIGW2}). But, one has to make the following changes: the coordinate time $t$ has to be replaced by conformal time $\tau$, and the expression for $h_+(\tau,z)$ is given by Eq. (\ref{eq:GWexp}). The resulting equations are as follows

\begin{align}
     \label{eq:typeIIGW1cos}
     \Box \tilde{\Phi} - 2 g^2_{\rm{YM}} l^2\; \bar{\Psi}\; \bar{\Phi}\; \tilde{\Psi} - g^2_{\rm{YM}} l^2 \; \bar{\Psi}^2 \; \tilde{\Phi} & = - \p_\mu h_+(\t,z) \p^\mu \bar{\Phi} - h_+(\t,z) (g_{\rm{YM}}^2 l^2\; \bar{\Phi}\; \bar{\Psi}^2)  \\ 
    \label{eq:typeIIGW2cos} 
    \Box \tilde{\Psi} - 2 g^2_{\rm{YM}} k^2 \bar{\Psi}\; \bar{\Phi}\; \tilde{\Phi} - g^2_{\rm{YM}} k^2\; \bar{\Phi}^2\; \tilde{\Psi} & = \p_\mu h_+(\t,z) \p^\mu \bar{\Psi} + h_+(\t,z) (g_{\rm{YM}}^2 k^2\; \bar{\Psi}\; \bar{\Phi}^2)
\end{align}
where $\bar{\Phi}$ and $\bar{\Psi}$ are the unperturbed YM wave solutions (\ref{eq:typeIIsol}) in de-Sitter background.

\begin{figure}[ht]
\begin{subfigure}{0.45\textwidth}
\includegraphics[width=0.8\linewidth,height=4cm]{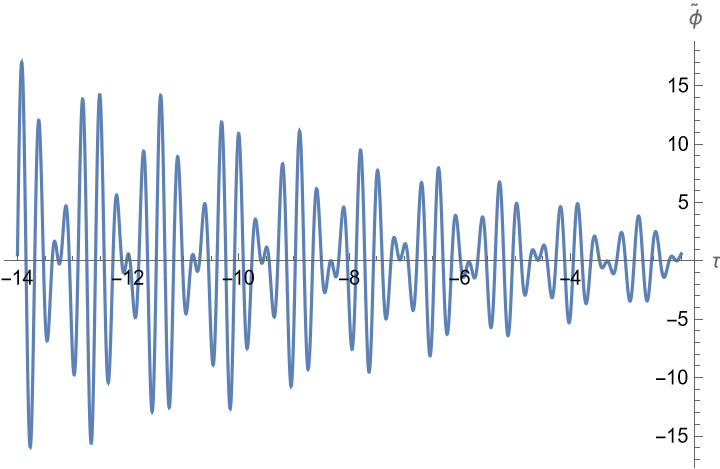}
\caption{$\tilde{\Phi}(1,\t)$}
\end{subfigure}
\begin{subfigure}{0.45\textwidth}
\includegraphics[width=0.8\linewidth, height=4cm]{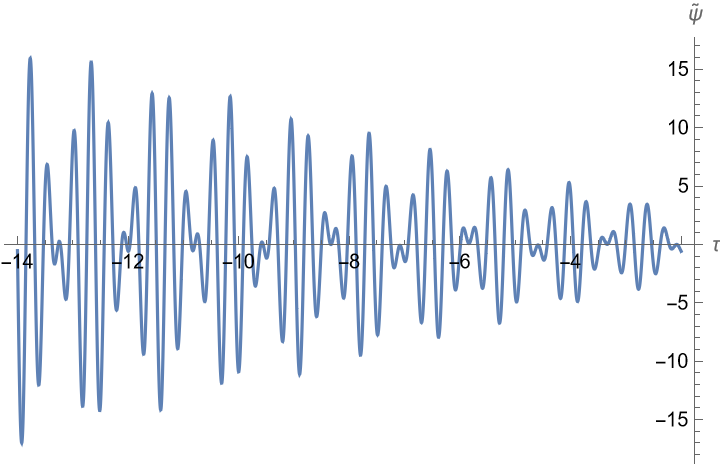}
\caption{$\tilde{\Psi}(1,\t)$}
\end{subfigure}
\caption{Figures showing $\tilde{\Phi}$ and $\tilde{\Psi}$ as a function of $\t$. We choose $z=1$, $l_1 = 1$, $k_2 =1$, $A_+ = 0.01$, $g_{\rm{YM}} = 0.5$ and $\omega_g=20$. }
\label{fig:perpsiphicosmo}
\end{figure}

We solve these equations numerically and the solutions are given in the following Fig. (\ref{fig:DTypeII}). We also plotted the behaviour of the perturbed functions as a function of time in Fig. (\ref{fig:perpsiphicosmo}).


\begin{figure}[ht]
\begin{subfigure}{0.45\textwidth}
\includegraphics[width=0.8\linewidth,height=4cm]{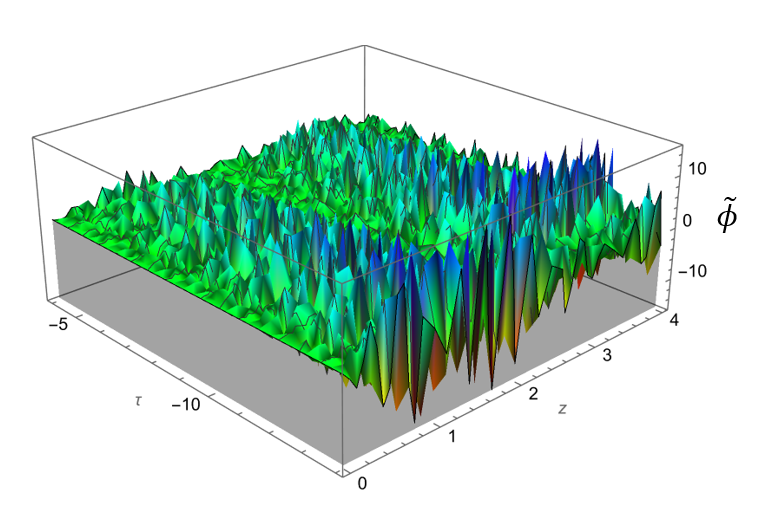}
\caption{$\tilde{\Phi}$}
\end{subfigure}
\begin{subfigure}{0.45\textwidth}
\includegraphics[width=0.8\linewidth, height=4cm]{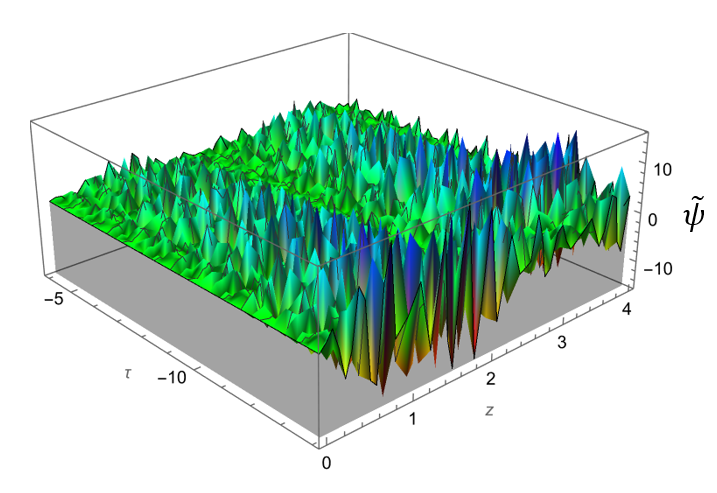}
\caption{$\tilde{\Psi}$}
\end{subfigure}
\caption{Figures showing $\tilde{\Phi}$ and $\tilde{\Psi}$ as a function of $\t$ and $z$. We choose $l_1 = 1$, $k_2 =1$, $A_+ = 0.01$, $g_{\rm{YM}} = 0.5$ and $\omega_g=20$. }
\label{fig:DTypeII}
\end{figure}

As one can see above, the sinusoidal waves are attenuated to perturbations due to the gravitational wave. The word attenuation is used in physics to describe the decrease in amplitude due to the propagation of a wave in a medium. Here the effect is similar as the initial wave changes or shows oscillations in amplitude due to the presence of a gravitational wave in a cosmological background. The amplitude of the perturbations are functions of the conformal time $\tau$ and the behaviour is similar to that observed in \cite{Akash} for Electromagnetic waves. The self-interaction of the waves does not change the overall pattern of the attenuation to a great extent from the Abelian perturbation as observed for the given value of $g_{\rm YM}$.

\subsection{Summary}
In this section, we obtain the YM fields in the background of conformal De-Sitter space and gravitational waves. We solve for the Type I, Coleman, and Type II waves (as defined in section [2]) in the background of gravitational waves in De-Sitter space.  For the Type I and Coleman waves, we obtain the solutions analytically and show that the GW wave influences the YM fields. For the Type II waves, the equations are solved numerically and we use the word attenuation to describe the plots in Figure (\ref{fig:perpsiphicosmo}), mainly as there is a dissipation effect due to the net `medium' of GW wave in De-sitter cosmology. Note our analysis of the YM waves in the De-Sitter space is also new. Our analysis is classical but paves the way for the quantum study of YM fields GW interactions in early Universe cosmology. Note as previously the amplitude of the perturbed Type II waves depend on the $g_{\rm YM}$. In addition to Type II waves, we find there is an interaction between YM wave and GW when propagating in the same direction for Type I waves too.
\section{Yang Mills waves and Gravitational Wave interaction in Electroweak Symmetry Broken phase}

In this section, we explicitly discuss the Electroweak broken phase gauge theory, before studying it in the GW background. The notations for the fields are thus redefined to make them consistent with the conventional $W,Z$ gauge boson symbols. The Electro-weak Higgs Lagrangian is \cite{pich}:
$$ \mathcal{L}_{EW} = -\frac14 W_{\mu \nu}^a W^{\mu \nu}_a - \frac14 B_{\mu \nu} B^{\mu \nu} + (D_{\mu} \phi)^{\dag} (D^{\mu} \phi) + V(|\phi|^2)$$
where $D_{\mu} \phi= \left(\partial_{\mu} - i \frac{g_{\rm YM}}{2} W_{\mu} - i \frac{g_{\rm Y}}{2} B_{\mu}\right) \phi$, 
and $W_{\mu}^a T^a$ is the SU(2) Gauge field, $T^a$ the Pauli matrices,  and $B_{\mu}$ is the hypercharge U(1) gauge field, $g_{\rm YM}$ and $g_{\rm Y}$ are the SU(2) and U(1) gauge coupling constants and $\phi$ is the scalar Higgs field. $W_{\mu \nu}^a= \partial_{\mu} W_{\nu}^a -\partial_{\nu} W_{\mu}^a + g_{\rm YM} \epsilon^{a b c} W_{\mu}^b W_{\nu}^c$ is the SU(2) field strength and $B_{\mu \nu}= \partial_{\mu} B_{\nu}- \partial_{\nu} B_{\mu}$ is the U(1) gauge field strength.
For Higgs acquiring a vacuum expectation value (vev), using the dynamics of the $V(\phi)$ potential, the $SU(2)\otimes U(1)$ symmetry is spontaneously broken to a new $U(1)$ gauge field obtained using electroweak rotation by an angle $\theta_W$. This rotation is implemented in the $W_{\mu}^3$ and the $B_{\mu}$ fields, the U(1) subgroup of SU(2), and the hypercharge U(1) field. Given the vev of the Higgs field, as 
$$\phi= \left(\begin{array}{c}0\\  \frac{v}{\sqrt{2}}\end{array}\right).$$

A mass term is generated for the Gauge fields from the $(D_\mu \phi)^{\dag}(D^{\mu} \phi)$, the Higgs field Kinetic term. This is of the form
$$ \frac{v^2}{4} g_{\rm YM}^2 W_{\mu}^{+} W^{\mu \  -} + \frac{g_{\rm Y}^2 + g_{\rm YM}^2}{8} v^2 \left(-\frac{g_{\rm YM}}{\sqrt{g_{\rm Y}^2+g_{\rm YM}^2} }W^3+ \frac{g_{\rm Y}}{\sqrt{g_{\rm YM}^2+g_{Y}^2}} B\right)^2 ,$$
where $ W_\mu^\pm = \frac{1}{\sqrt{2}}(W_\mu^1 \pm W_\mu^2)$ and using $$\cos\theta_W= \frac{g_{\rm YM}}{\sqrt{g_{\rm Y}^2+g_{\rm YM}^2}},  \   \   \   \   \   \   \  \sin\theta_W= \frac{g_{\rm Y}}{\sqrt{g_{\rm Y}^2+g_{\rm YM}^2}}, $$ the new fields generated are of the following form.  
$$A_{\mu}= \sin\theta_W W^3_{\mu} + \cos\theta_W B_{\mu};  \  \   \   \  \   Z_{\mu}= \cos\theta_W W^3_{\mu}-\sin\theta_W B_{\mu}.$$
The $A_{\mu}$ field is identified with the Electromagnetic field of today and $Z_{\mu}$ is the massive gauge boson, remnant of the full SU(2) symmetry. 
If we see the Kinetic term for the Gauge fields under this redefinition (Electroweak rotation), one gets:
$$-\frac{1}{4} \left[ (\partial_{\mu} W_{\nu}^a-\partial_{\nu} W_{\mu}^a)^2 + 2 g_{\rm YM} \epsilon_{a b' c'} W^{\mu b'} W^{\nu c'} (\partial_{\mu} W^a_{\nu}-\partial_{\nu} W^a_{\mu}) + g_{\rm YM}^2 ((W_{\mu}^aW^{\mu}_a)^2 - W^a_{\mu}W_{\nu a} W^{b \mu} W_{b}^{\nu})\right]-\frac14 B_{\mu \nu} B^{\mu \nu}$$
Using the Oh-Teh Ansatz \cite{OT}, $W^1_{\mu}= k_{\mu} \Phi$ and $Z_{\mu}= l_{\mu} \Psi $, (the a=1,3 SU(2) components are non-zero) one gets the two differential equations from the total Lagrangian as
\bea
l^2 \Box \Psi- (\vec{l}\cdot \vec{\partial}) \Psi - g_{\rm YM}^2 k^2 l^2 \Psi \Phi^2 \cos^2\theta_W + \frac{v^2}{4} (g_{\rm YM}^2 +g_{\rm{Y}}^2) l^2 \Psi & =& 0 \\
k^2 \Box \Phi- (\vec{k}\cdot \vec{\partial}) \Phi - g_{\rm YM}^2 k^2 l^2 \Phi \Psi^2 \cos^2\theta_W + \frac{v^2}{4} g_{\rm YM}^2 k^2 \Phi & =& 0. 
\eea
We assume as previously, the same formula for the $l_{\mu}$ and $k_{\mu}$ such that $l_\mu \p^\mu \Psi = k_\mu \p^\mu \Phi = 0$. To obtain an analytic solution, we set $\Phi=\Psi$. Given that $e=g_{\rm YM} \sin\theta_W= g_{\rm Y}\cos\theta_w$, where $e$ is the electronic charge, and the $W$ Boson mass is $80 ~{\rm GeV/c^2}$ and $Z$ boson is $91~ {\rm GeV/c^2}$, the approximation of setting the two fields as equal can be taken as zeroeth order in the $g_{\rm Y}$ constant.
The solution is given as 
\be
\Psi (z,t)=-\frac{v} {2 \cos\theta_W} \tanh\left(A \ z -\frac{\sqrt{4 A^2 - g_{\rm YM}^2 v^2/2}}{2} \ t +B\right)
\label{eqn:ans}
\ee
where $A, B$  are arbitrary constants. 
Similarly, the solution for $\Phi$ can be obtained accordingly. 

\begin{figure}[ht]
\begin{subfigure}{0.45\textwidth}
\includegraphics[width=0.8\linewidth,height=5.5cm]{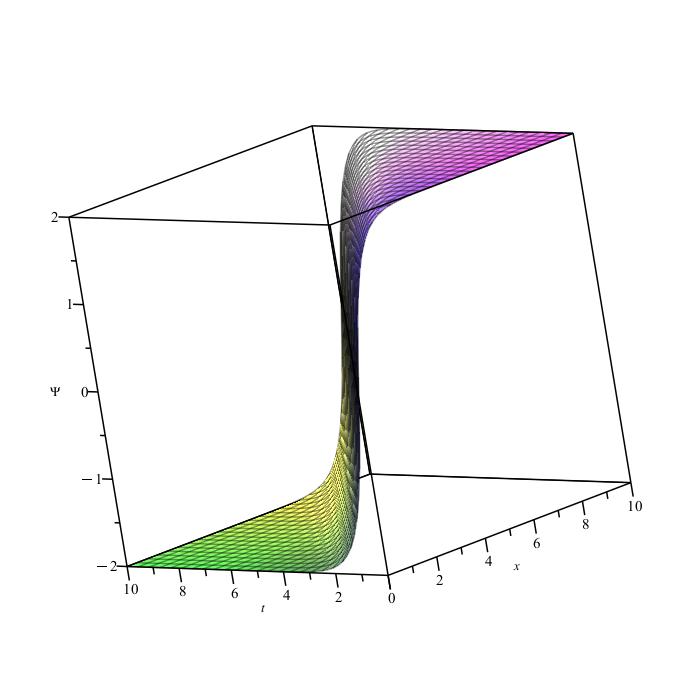}
\caption{The exponential YM gauge field with mass.}
\label{fig:poynt}
\end{subfigure}
\begin{subfigure}{0.45\textwidth}
\includegraphics[width=0.8\linewidth, height=5.5cm]{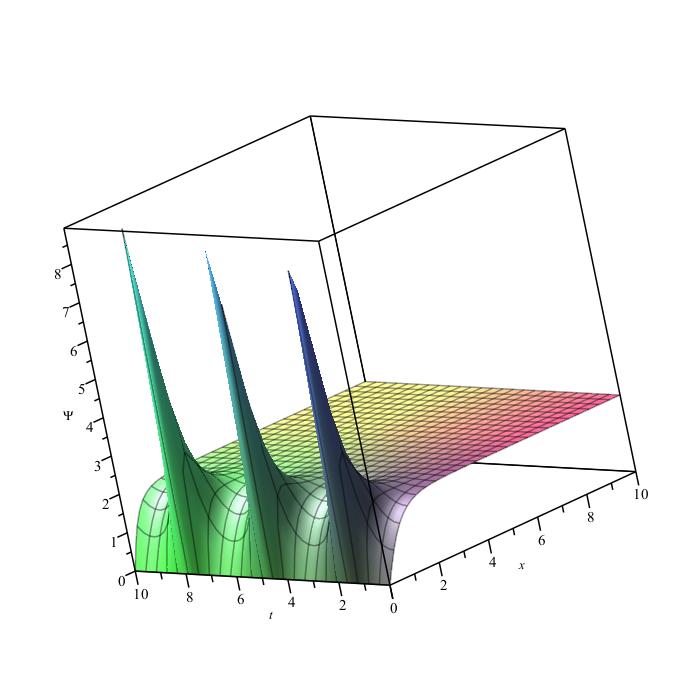}
\caption{Oscillatory YM gauge field with mass.}
\label{fig:poynt1}
\end{subfigure}
\caption{Figures showing $\Psi$ as a function of $z$ and $t$. }
\end{figure}

The plots of these show exponential behaviour, depending on the constants, and boundary conditions. If the constants $4 A^2-g_{\rm YM}^2 v^2/2 >0$, the function is exponential (Fig. (\ref{fig:poynt})), and if $4 A^2-g_{\rm YM}^2 v^2/2<0$, the function shows oscillatory behaviour (Fig. (\ref{fig:poynt1})). 

\subsection{Interaction with Gravitational wave}
\begin{figure}[ht]
    \centering
    \includegraphics[width=0.5\linewidth, height=6cm]{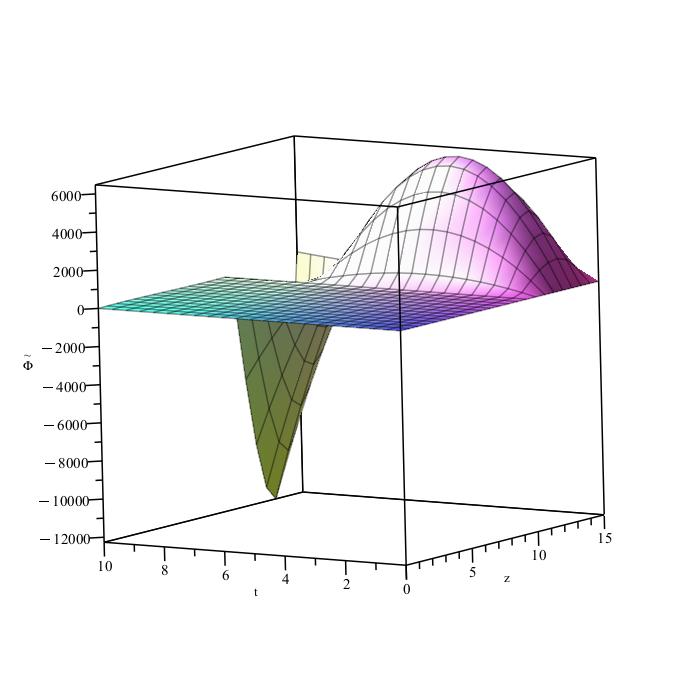}
    \caption{Perturbed function ($\tilde{\Phi}$) or perturbation of massive YM field due to GW grows rapidly. We choose $A=4$, $B=0$, $g^2_{\rm YM} \cos^2\theta_W =0.1,  v^2 g^2_{\rm YM}=40\; \textrm{and}\; \omega_g=1$. }
    \label{fig:poynt2}
\end{figure}

If one uses the metric as $g_{\mu \nu}= \eta_{\mu \nu} + h_{\mu \nu}$, one gets the equation of motion from the same Lagrangian as
\bea
&&k^2\Box \Phi- (\vec{k}\cdot \vec{\partial})^2 \Phi - g_{\rm YM}^2 \cos^2\theta_W \Psi^2 \Phi l^2 k^2 + \frac{v^2}{4} g_{\rm YM}^2 k^2 \Phi = \partial_{\mu} (h^{\nu \rho})(k_{\nu}k_{\rho} \partial^{\mu} \Phi - k_{\nu} k^{\mu} \partial_{\rho} \Phi) \nonumber \\ && + k^2 h^{\mu \lambda} \partial_{\mu} \partial_{\lambda} \Phi - h^{\mu \lambda} k^{\rho} k_{\lambda} \partial_{\mu} \partial_{\rho} \Phi - h^{\nu \rho} k_{\nu} k_{\rho} \Box \Phi - k_{\nu} h^{\nu \rho} \partial_{\rho} (\vec{k}\cdot \vec{\partial}) \Phi 
\eea  
A similar equation for $\Psi$ is obtained. In the Linearized approximation, we take $\Phi=\bar{\Phi} + \tilde\Phi$ and $\Psi= \bar{\Psi} + \tilde{\Psi}$. In the previous discussion, we did not have to take the specific choice of the direction of $k_{\mu}$ and $l_\mu$. Here in the following, we take the $l_{\mu}=\delta_{\mu}^1$, and $k_{\mu}=\delta_{\mu}^2$, and keeping the linear terms in perturbation one has
\bea
&& \Box \tilde{\Phi} - 2 g_{\rm YM}^2 \cos^2\theta_W  \; \bar{\Psi}\; \bar{\Phi}\; \tilde{\Psi}  - g_{\rm YM}^2 \cos^2\theta_W \bar{\Psi}^2 \tilde{\Phi} + \frac{v^2}{4} g_{\rm YM}^2  \tilde{\Phi}= - \partial_{\mu} h^{22} \partial^{\mu} \bar{\Phi} \nn \\ && - h^{22} \Box \bar{\Phi} - \frac{v^2}{4} g_{\rm YM}^2 h^{22} \bar{\Phi}
\eea
As previously, to facilitate a solution, we set the $\bar{\Psi}=\bar{\Phi}$ and $\tilde\Phi=\tilde\Psi$. This makes the equation solvable, though we use numerical methods. We choose the exponential solution for unperturbed functions. The solution obtained is oscillatory, and the amplitude increases with time. However, what is crucial is that, unlike the previous discussions, the gravitational wave amplitude is multiplied by the W, Z Boson masses, which in natural units is $80 \ {\rm Gev} \approx 4.06 \times 10^{17}/{\rm m}$. This makes the perturbations acquire order 1 magnitude very quickly (Fig. (\ref{fig:poynt2})). 
Thus the perturbation theory breaks down, and one has to solve for the exact solution of the differential equation.

\begin{figure}[ht]
    \centering
    \includegraphics[width=0.5\linewidth, height=6cm]{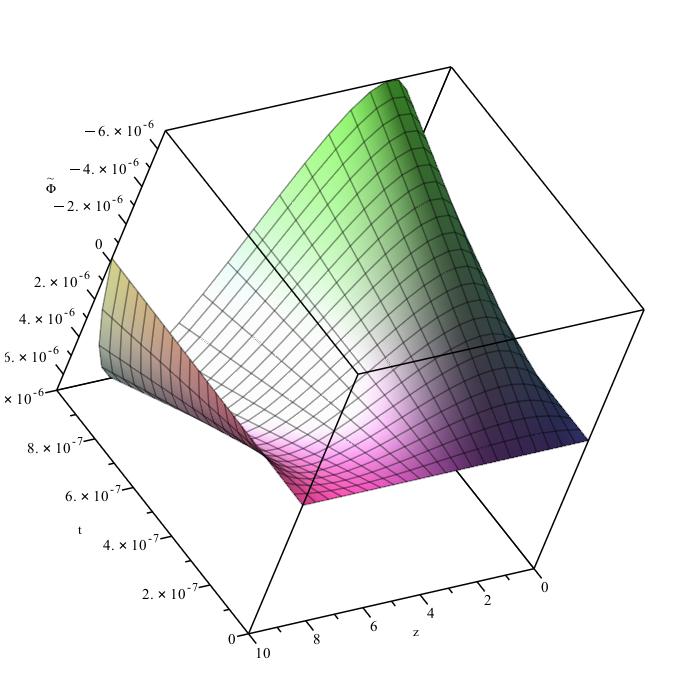}
    \caption{Perturbation ($\tilde{\Phi}$) or perturbation of massive YM field due to GW, plotted for short time $t< 10^{-6}$, $g^2_{\rm YM} \cos^2\theta_W =0.1, v^2 g^2_{\rm YM}=40\; \rm{and}\;  \omega_g=1$. In this short time, the perturbation is in control.}
    \label{fig:poynt3}
\end{figure}

However, due to the mass term, the W, Z particles decay very quickly. A typical W particle has a half-life of about $10^{-25} s$, keeping that in mind, we make a representative plot of time $0<t<10^{-6} s$ with $v^2g_{\rm YM}^2 = 40$ and $g_{\rm{YM}}^2 \cos^2\theta_W=0.1$. The gravitational wave is taken of the angular frequency of $1 {\rm Hz}$ and amplitude $10^{-2}$ amplitude. The constants in Eq. (\ref{eqn:ans}) are set to $A=4, B=0$.  The perturbation is infinitesimal in that range, and it also shows the induced oscillations due to the inhomogeneous term in the equation (Fig. (\ref{fig:poynt3})). Note the constants have been chosen such that the behaviour of the massive gauge field is hyperbolic and not oscillatory to obtain the solution.

As the gravitational wave is a solution to linearized Einstein equations, the perturbation analysis is the correct approximation to the system of equations discussed above. To study the non-perturbative regime, one has to obtain the complete solution to the Einstein-Yang-Mills system. An extension of our work to predict the behaviour of W, Z Bosons in presence of gravitational waves on Earth, would require quantum scattering calculations and collider physics details. This is  work in progress.

\subsection{Summary}
In this section, we have analyzed the Electro-weak YM fields coupled with a Higgs. As the Higgs Boson has a vacuum expectation value, the YM SU(2) symmetry is broken. However, using techniques similar to the previous two sections, we solved the massive YM fields using the Type II ansatz. Note that our solution in flat space-time Eq. (\ref{eqn:ans}) is new, and should be useful for Nuclear Physics and Particle physics purposes.  Further, we solve for the perturbation induced by the GW on these YM fields, and we find that the numerical solution grows very rapidly out of the linear regime. However, if we observe the perturbation for a short time, the solution is valid. These are classical solutions, but provide the input for studying the quantum mechanics of $W, Z$ bosons in presence of GW.

\section{Conclusion}
In this paper, we have studied the interaction of the GWs with YM gauge fields. Our calculation is based on the physics of a system where the YM field is `impacted' with a GW, and the YM field is perturbed. The perturbation can be detected as in an EM wave \cite{Akash}. We find that the various types of progressive YM plane waves identified as Type I (no self-interaction) and Type II (with self-interaction) were perturbed by the GWs. To facilitate a realistic experiment on earth, we also calculate the same interaction in the broken phase of the YM field. For the Type I fields, the perturbations induced by the GW waves are similar to the EM wave perturbations studied in \cite{Akash}. The YM waves which are progressive in time and space deviate from their direction. This we have demonstrated graphically. In the study of Type II perturbations, where the self-interaction terms of the YM fields are non-zero, the propagation of the perturbations has been analyzed using spectral analysis. The dependence on the YM coupling constant $g_{\rm YM}$ is through the amplitude of the perturbed waves which increases with weaker interactions. This is consistent with the Jacobi function solutions of the unperturbed YM solutions. In the study of the massive W, Z Boson fields, we find that in the very short time of the existence of the W, Z boson, the GW perturbs the field, and this can be obtained numerically. However, the non-perturbative discussion for this will likely provide more insight into the physics of the system. In particular, we can ask the question of whether the presence of a GW serves to increase the `mass' of the W, Z Boson, and therefore change its decay pattern in a collider? Our results will also be relevant for the heavy ion collisions producing de-confined quarks and gluons when extended to the SU(3) YM fields \cite{hev}. As in EM theory, where we are suggesting ways to detect GW using MASERS etc, we expect new physics of detection of both GW and YM fields to emerge from this research program. Whereas we are not expecting a YM laser gun in the near future, our aim is to show that there are experiments with W, Z Bosons observations in the broken phase today, and in the unbroken phase in the early Universe, and these react to GW. We know/obtain their behaviour using YM field equations, and show how these are influenced by GW in this paper. In the future, our program will facilitate the comparison of scattering amplitudes of the gravitational wave with electroweak Bosons in the collider. We hope to extend our calculations to SU(3) YM fields and gluons. Our work will also shed light on one-loop renormalization of the theories \cite{huan}. In particular the study of propagating YM waves interactions with GW will shed light on quark gluon plasma of the early universe. The study of GW effect on YM condensates and plasma waves is going to appear in the near future \cite{condensate}.
\vspace{0.5cm}

\noindent
{\bf Acknowledgement:} NRG would like to thank MITACS for accelerator grant.

\noindent
{\bf Declarations:}
The authors have no competing interests to declare that are relevant to the content of this article. They also declare no conflict of interest with anyone or the funding agencies for conceptualizing and the publication of the results. The authors declare that the results/data/figures have not been published elsewhere, nor are they in consideration for publication by another publisher. The authors declare that the data supporting the findings of this study are available within the paper.


\printbibliography

\end{document}